\documentclass[twocolumn]{autart}    

\usepackage{graphicx}          

\usepackage{enumerate}
\usepackage{subfigure}
\usepackage{multirow}
\usepackage{mathrsfs}
\usepackage{amsfonts}
\usepackage{amsmath}
\usepackage{amssymb}

\theoremstyle{plain}
\newtheorem{coro}[thm]{Corollary}
\newtheorem{assumption}[thm]{Assumption}

\begin{document}

\begin{frontmatter}

\title{Modelling and Filtering for Non-Markovian Quantum Systems\thanksref{footnoteinfo}} 

\thanks[footnoteinfo]{This paper was not presented at any IFAC
meeting. This research was supported under Australian Research Councils
Discovery Projects and Laureate Fellowships funding schemes (Projects
DP140101779 and FL110100020), the Chinese Academy of Sciences President's
International Fellowship Initiative (No. 2015DT006), and the Air Force Office of Scientific Research (AFOSR) under agreement FA2386-16-1-4065. Corresponding author
S.~Xue,  Tel. : (+61) 02-6268-9461}

\author[adfa]{Shibei Xue}\ead{xueshibei@gmail.com},    
\author[anu]{Thien Nguyen}\ead{thien.nguyen@anu.edu.au},               
\author[anu]{Matthew~R.~James}\ead{Matthew.james@anu.edu.au},  
\author[google]{Alireza Shabani}\ead{shabani@google.com},
\author[adfa]{Valery Ugrinovskii}\ead{v.ugrinovskii@gmail.com},
\author[anu]{Ian~R.~Petersen}\ead{i.r.petersen@gmail.com}

\address[adfa]{School of Information Technology and Electrical Engineering, University of New South Wales at the Australian Defence Force Academy, Canberra, ACT 2600, Australia}  
\address[anu]{Research School of Engineering, Australian National University, Canberra, ACT 2600, Australia}             
\address[google]{Quantum Artificial Intelligence Lab, Google, 340 Main St. Venice, CA 90291, U.S.A.}        

\begin{keyword}                           
Non-Markovian quantum systems; Quantum stochastic differential equation; Whitening quantum filter; Quantum Kalman filter; Quantum colored noise.                     
\end{keyword}                             

\begin{abstract}                          
This paper presents an augmented Markovian system model for non-Markovian
quantum systems. In this augmented system model, ancillary systems are
introduced to play the role of internal modes of the non-Markovian
environment converting white noise to colored noise. Consequently, non-Markovian
dynamics are represented as resulting from direct interaction of the
principal system with the ancillary system. To demonstrate the utility of
the proposed augmented system model, it is applied to design  whitening
quantum filters for non-Markovian quantum systems. Examples are presented
to illustrate how whitening quantum filters can be utilized for estimating
non-Markovian linear quantum systems and qubit systems. In particular, we
showed that the augmented Markovian formulation can be used to
theoretically  model the environment for an observed non-Markovian behavior
in a recent experiment on quantum dots~\cite{Frey}.
\end{abstract}

\end{frontmatter}

\section{Introduction}
Control of open quantum systems has been rapidly advancing quantum information technology in recent years~\cite{Breuer,Mirrahimi20051987,Ticozzi20092002,Kuang200898,Qi2010333,6189045,Dong2012725,Pan2016147}, where an open quantum system refers to a quantum system interacting with an environment or other quantum systems. 

Among open quantum systems, the most widely investigated class of quantum systems is that comprising quantum systems coupled with memoryless environment. Evolution of such systems
 can be described by master equations~\cite{Breuer} in the Schr{\"o}dinger
 picture and Langevin equaitons~\cite{Breuer} or quantum
 stochastic differential equations~\cite{bouten} in the Heisenberg
 picture. Both mathematical models give rise to Markovian dynamics, thus
 this class of open quantum systems is commonly referred to as Markovian
 quantum systems.
In addition, Markovian quantum systems can be coupled to a field satisfying a singular commutation relation, e.g., quantum white noise~\cite{Gardiner1985}. In terms of its stochastic description, the field has an independent increment over an infinitesimal time interval, which satisfies a non-demolition condition and can be measured to continuously extract information of the quantum system~\cite{Belavkin}.
These properties of the environment are analogous to properties of a classical white noise and have served as a foundation for the well established quantum filtering theory aimed at estimation of Markovian quantum systems~\cite{bouten}. Such a theory underpins a number of successful control applications such the design of
real-time feedback control laws for cooling a quantum particle~\cite{Doherty1999}, and stabilizing states~\cite{Mirrahimi} or entanglement~\cite{Yamamoto2007981,5395610} of a quantum system.

However, many problems of interest involve more complicated environmental
influences, which cannot be handled within the Markovian setting and
require treating the environment as quantum colored noise. This
necessitates the investigation of non-Markovian behavior of quantum
systems~\cite{Tan2011,Xue2011,PhysRevA.58.1699,XuePRA2012,Barch2012PRA,Xue2015,Xue2016}.
A non-Markovian quantum system is a quantum system interacting with an
environment with memory effects.  To describe  dynamics of non-Markovian
quantum systems involving quantum colored noise, several models have been
developed, for example, non-Markovian Langevin equations where
non-Markovian effects are embedded in a memory kernel
function~\cite{Tan2011} and time-convolutionless master equations where a
time-varying damping function characterizes the non-Markovian damping
processes~\cite{Breuer}, etc. However, these existing models for
non-Markovian quantum systems are not compatible with the quantum filtering
theory. In addition, unlike the quantum white nose,
the quantum colored noise does not satisfy the singular commutation
relations. For that reason, non-Markovian models are difficult to use for
processing quantum measurements. Once the quantum colored noise is
measured, the states of the quantum system interacting with this noise will
be demolished.

A standard approach in classical control systems analysis and design is
whitening of the colored noise by introducing additional dynamics so as to
express the system with non-Markovian effects of colored noise as an
augmented system model governed by a white noise. Thus a filter for the
system involving colored noise can be  constructed~\cite{KS72}; such filter
is often referred to as whitening filter. Similar ideas have been explored for
quantum systems as well. A pseudo-mode method was proposed for effectively
simulating non-Markovian effects by using a Monte Carlo
wave-function~\cite{PhysRevA.50.3650,PhysRevA.80.012104}, which was applied
to model the energy transfer process in photosynthetic
complexes~\cite{JCP}.
Also, the dynamics of non-Markovian quantum systems can be described by
using a hierarchy equation approach~\cite{PhysRevA.85.062323}, which has
been applied to indirect measurement of a non-Markovian quantum
system~\cite{shabani2014}. An augmented system approach has been applied to
obtain a quantum filter for quantum systems interacting with non-classical
fields using a field-mediated connection method in a situation where the
non-Markovian system does not introduce backaction on the
environment~\cite{gough2012PRA}.

In this paper we present a systematic augmented Markovian system approach to
modelling non-Markovian quantum systems. To capture effects of the
non-Markovian environment, we introduce ancillary systems to augment a
principal system of interest, which are realized by linear
open quantum systems. Compared to the principal system, the augmented system
model is defined on an augmented Hilbert space. 
Also, we introduce a spectral factorization method to determine the
structure of linear ancillary systems to ensure that its fictitious output has a
power spectral density which is identical to that of the non-Markovian
environment under consideration. Nevertheless, while these elements of our
model follow the classical system modelling, the proposed model has
a distinctively quantum feature in that the quantum plant and its
non-Markovian environment mutually influence each other. This feature
distinguishes quantum system-environment interactions from the classical
case where the classical colored noise disturbs a plant but not \emph{vice
  versa}~\cite{ORE}. To account for this special feature of non-Markovian
quantum systems, in the proposed model the ancillary system is coupled to
the principal system via their direct interactions rather than the
field-mediated connection; cf.~\cite{gough2012PRA}.




To describe the augmented Markovian system model for the non-Markovian
quantum system, the paper adopts so-called $(S,L,\mathsf{H})$ description,
where the internal energy, the couplings to the environment and the
scattering process of the environmental field for a quantum system are
captured by a Hamiltonian $\mathsf{H}$, a coupling operator $L$, and a
scattering matrix $S$, respectively.
An advantage of this approach is that it allows to describe system-environment
interactions systematically using the formalism of quantum
stochastic differential equations. To demonstrate this advantage and
the utility of the proposed augmented Markovian modelling of
non-Markovian quantum dynamics, we show how the proposed approach can be
utilized to obtain whitening quantum filters for non-Markovian linear quantum
systems and qubit systems. As an example application, this augmented
Markovian model is utilized to explore quantum colored noise in a recent
experiment for a hybrid solid-state quantum system~\cite{Frey}.

The paper is organized as follows. In section~\ref{sec2}, we briefly review
a general description of Markovian quantum systems. Based on this
description, an augmented Markovian system model is presented for
non-Markovian quantum systems in section~\ref{sec3}, where a spectral
factorization method is proposed to obtain an ancillary linear quantum
system model for a quantum environment with a given spectrum. Next, the
application to derivation of a whitening quantum filter is presented in
section~\ref{sec4} where examples of filters for linear non-Markovian
quantum systems and single qubit systems are obtained. In section~\ref{sec5}, an
augmented Markovian system model is utilized to obtain an improved model
for an experiment involving a hybrid solid-state system. Finally,
conclusions and discussions are given in Section~\ref{sec6}.

\section{A brief review of the Markovian quantum system model}\label{sec2}

In this section, we will briefly introduce some standard facts about
Markovian quantum systems. For more details, we refer the reader to
references~\cite{Gardiner,bouten}.
\subsection{Quantum white noise}
In quantum physics, it is customary to describe an environment field as the
Fourier transform of the annihilation operator of the field acting on the
so-called Fock space $\mathfrak{F}$~\cite{Gardiner}; that is
\begin{equation}\label{0}
  b(t)=\frac{1}{\sqrt{2\pi}}\int_{-\infty}^{+\infty}b(\omega)e^{-{\rm
      i}\omega t}{\rm d}\omega,
\end{equation}
The operator on the Fock space defined by (\ref{0}) is called the quantum
white noise field operator. It satisfies singular commutation relations
\begin{equation}\label{0-1}
  [b(t),b^\dagger(t')]=\delta(t-t'),\quad [b(t),b(t')]=0,
\end{equation}
where $[\cdot,\cdot]$ is the commutator of operators, i.e., $[f,g]=fg-gf$
for two operators $f$ and $g$ with suitable domain and image spaces. The
symbol $^\dagger$ denotes the complex conjugate of an operator.

The quantum white noise field can be interpreted as a quantum stochastic
process. Associated with the field $b(t)$ are an integrated operator
$B(t)=\int_{t_0}^t b(\tau){\rm d}\tau$, known as
$B(t)$ is a quantum Wiener process, and
its adjoint $B^\dagger(t)=\int_{t_0}^t b^\dagger(\tau){\rm d}\tau$. They
satisfy the commutation relations~\cite{bouten}
\[
[B(t),B^\dagger(\tau)]=\min(t,\tau), \quad
[B(t),B(\tau)]=0.
\]
Under a common assumption that the initial state of the field is a vacuum field, the process $B(t)$ is analogous to the standard Wiener process and the process $b(t)$ is analogous to a Gaussian white noise with zero mean. 
For convenience, we summarize the Ito rules
for the quantum infinitesimal increments of $B$, $B^\dagger$ in a vacuum
state~\cite{Gardiner} which will be used in subsequent calculations,
\begin{eqnarray}\label{0-2}
  {\rm d}B(t){\rm d}B^\dagger(t)={\rm d}t,~~~~~~&&~~{\rm d}B^\dagger(t){\rm d}B(t)=0,\nonumber\\
  {\rm d}B(t){\rm d}B(t)=0,~~~~~~&&~~{\rm d}B^\dagger(t){\rm d}B^\dagger(t)=0.
\end{eqnarray}


\subsection{Quantum stochastic differential equation}
\begin{figure}
  \includegraphics[width=8.5cm]{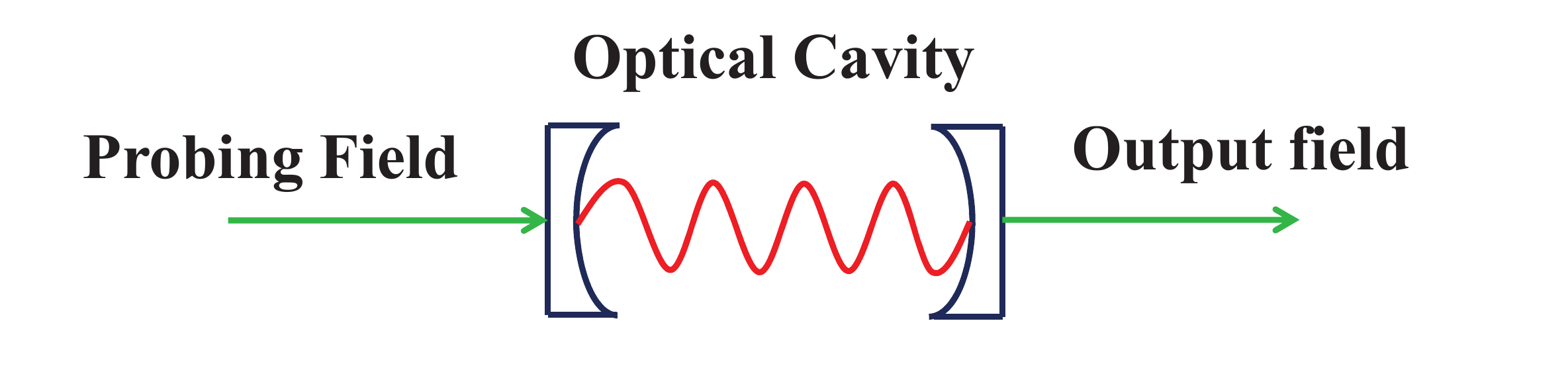}\\
  \caption{A standard Markovian quantum system model, i.e., optical modes trapped in a cavity
probed by a white noise field.}\label{f0}
\end{figure}

Markovian quantum systems have been extensively investigated since they are
suitable models for many physical systems. For example, optical modes
trapped in a cavity (see Fig.~\ref{f0}) probed by a white noise field
exhibit Markovian dynamics.

Markovian quantum systems can be described by
quantum stochastic differential equations.
Consider a Markovian quantum system and the associated triple
$(S,L,\mathsf{H})$ characterizing its scattering matrix, the coupling
operator and the Hamiltonian, respectively. In quantum mechanics, both the
coupling operator and the Hamiltonian are represented as operators mapping
an underlying Hilbert space $\mathfrak{h}$ into itself, As for the
scattering matrix $S$, from now on, we will assume $S={\rm I}$ since for
simplicity the scattering process will not be considered here.

The unitary evolution operator $\Theta(t)$ of this
Markovian quantum system is defined on the tensor product Hilbert space
$\mathfrak{h}\otimes\mathfrak{F}$, it is known to satisfy the following
quantum It$\rm
\bar{o}$ stochastic differential equation
\begin{equation}\label{2}
  {\rm d}\Theta(t)={\Big (}-{\big(}{\rm i}\mathsf{H}+\frac{1}{2}L^\dagger
  L{\big)}{\rm d}t+{\rm d}B^\dagger(t) L-L^\dagger {\rm d}B(t){\Big
    )}\Theta(t).
\end{equation}
In the Heisenberg picture, all quantum mechanical quantities of interest
affected by the system evolution are operators that initially defined on
the Hilbert space $\mathfrak{h}$, and evolve on the Hilbert space
$\mathfrak{h}\otimes\mathfrak{F}$. Specifically, an arbitrary operator $X$
defined on $\mathfrak{h}$ gives rise to the evolution $j_t(X)$ of operators
on the Hilbert space $\mathfrak{h}\otimes\mathfrak{F}$ defined as
$j_t(X)=\Theta^\dagger(t) X\Theta(t)$. From (\ref{2}), it follows that
this evolution satisfies the quantum stochastic differential equation
\begin{equation}\label{3}
 {\rm d}j_t(X)=j_t(\mathcal{G}(X)){\rm d}t+{\rm
   d}B^\dagger(t)j_t([X,L])+j_t([L^\dagger,X]){\rm d}B(t).
\end{equation}
Here $\mathcal{G}$ refers to the \emph{Lindblad generator}~\cite{bouten}
\begin{equation}\label{4}
\mathcal{G}(X)=-{\rm i}[X,\mathsf{H}]+\mathcal{L}_L(X),
\end{equation}
and the notation $\mathcal{L}_{L}(X)$ refers to the \emph{Lindblad
  superoperator} defined as
\[
\mathcal{L}_{L}(X)=\frac{1}{2}L^\dagger[X,L]+\frac{1}{2}[L^\dagger,X]L.
\]
Quantum stochastic differential equations have been widely used in the
analysis and control of Markovian quantum
systems~\cite{bouten,hinfinity,NYamamoto,maalouf}.

\subsection{Input-output relations}\label{inputoutput}
When a quantum white noise field passes through a quantum object and
interacts with it, the resulting field is called an output field. It can be
observed via measurement, e.g., homodyne
detection~\cite{Gardiner}. Mathematically, the output field is described as
 $ B_{\rm out}(t)=\Theta^\dagger(t) B(t)\Theta(t)$. The quantum
infinitesimal increment for the output field can be written as a quantum
stochastic differential equation:
\begin{equation}\label{5}
  {\rm d}B_{\rm out}(t)=j_t(L){\rm d}t+{\rm d}B(t),
\end{equation}
which shows that the output field not only carries information about the
quantum object but is also affected by the input noise. As a result, the
output field can be utilized in estimating the dynamics of the quantum
object~\cite{bouten,Yanagisawa2003,guofeng2011}.

\subsection{Master equation}\label{MS}
While this paper is exclusively setup in the Heisenberg picture, it is
worth reminding about an alternative formulation known as the
Schr{\"o}dinger picture. In the Schr{\"o}dinger picture,
the state of a quantum system can be described by a wave function
$|\psi\rangle$ which is a complex vector in the Hilbert space
$\mathfrak{h}$. Based on the wave function $|\psi\rangle$, a trace class
operator $\rho=|\psi\rangle\langle\psi|$ on $\mathfrak{h}$ is defined known
as the density matrix of the quantum system~\cite{Yanagisawa2003}.

In contrast to the Heisenberg picture introduced in the previous
subsections, in which the operators evolve in time
and the states are time independent, the states of a quantum system evolve
in time while the operators are time independent in the Schr{\"o}dinger
picture. Of course, these two alternative representations of the quantum
evolution are equivalent~\cite{Breuer,Yanagisawa2003}.

The density matrix $\rho$ of a Markovian quantum system characterized by a
triple $(S,L,\mathsf{H})$ and interacting with a quantum white noise obeys
a so-called master equation
\begin{equation}\label{5-3}
 \dot \rho(t)=-{\rm i}[\mathsf{H},\rho(t)]+\mathcal{L}^*_L(\rho(t)),
\end{equation}
where the adjoint of the Lindblad superopertor $\mathcal{L}^*_L$
is calculated as $
\mathcal{L}^*_L(\rho(t))=\frac{1}{2}[L\rho(t),L^\dagger]+\frac{1}{2}[L,\rho(t)L^\dagger]$.

\section{An augmented model for non-Markovian systems}~\label{sec3}

\subsection{Preliminary remarks}

A non-Markovian quantum system is a quantum system in an environment with
memory, and is typically regarded as a quantum system disturbed by a quantum
colored noise. In this paper we propose to model such environments as
ancillary quantum systems excited by quantum white noise fields. This
approach resembles the classical approach to modelling colored noise
signals using shaping filters where the shape of the spectrum is related to
internal modes of the filter. Similarly, in this paper we assume that the
internal modes of the ancillary system correspond to dynamics of the
non-Markovian environment. However different from the classical case,
we show that the non-Markovian system behaviour and the system backaction on the
environment can be explained using a direct coupling mechanism of
environment-system interactions, within an augmented Markovian system
picture.

This section begins with introducing a general framework for constructing
such an augmented Markovian system model in Section~\ref{subs31}. Next in
Section~\ref{subs32}, we show that the class of proposed augmented models
is sufficiently rich in a sense that, given an environment with a rational
power spectral density of its quantum colored noise, an ancillary
\emph{linear} quantum system model for this environment can be constructed
to match that spectrum. This
implies that a large class of non-Markovian environments can be modelled
(exactly or with a sufficient accuracy) in terms of linear quantum systems
consisting of quantum harmonic oscillators.

\subsection{The general augmented Markov system framework}\label{subs31}

\subsubsection{The $(S,L,\mathsf{H})$ description of the augmented Markovian
  system model}

Consider a quantum system operating in a non-Markovian environment. This
system, which we call the principal system, has the following
$(S,L,\mathsf{H})$ description
\begin{equation}\label{8}
  \textsf{G}_p=({\rm I},L_p,\mathsf{H}_p),
\end{equation}
where $\mathsf{H}_p$ is the Hamiltonian of the principal system and $L_p$
is the coupling operator vector of the principal system with respect to a
probing field defined on a Fock space $\mathfrak{F}_p$. The operators
$\mathsf{H}_p$ and $L_p$ are defined on a system's Hilbert space
$\mathfrak{h}_p$. Note that the coupling operator $L_p$ does not describe
how the system is coupled with the environment, but allows the principal
system to be probed for measurement, by shining an input field through
probing channels and observing the output of the probing field so as to
construct quantum filters.


The principal quantum system interacts with its environment by exchanging
energy with the environment, i.e., they mutually influence each other. This
energy exchange is captured by a direct interaction Hamiltonian
\begin{equation}\label{9}
  \mathsf{H}_{pa}={\rm i}(c^\dagger z-z^\dagger c),
\end{equation}
where the operator vector $c$ describes the environmental effect, and $z$
is a coupling operator vector $z$ defined on the Hilbert space
$\mathfrak{h}_p$ of the principal system.

In what follows internal modes of the environment are assumed to be
stationary. Hence the quantum correlation matrix
$\langle c(t+\tau)c^\dagger (t)\rangle $ is independent of $t$; here
$\langle\cdot\rangle$ denotes the quantum expectation defined as
$\langle\cdot\rangle={\rm tr}[\cdot \rho_a]$, where $\rho_a$ is the initial
density matrix of the environment. Also the
power spectral density characteristics of the environment can be defined
in a standard fashion, as the Fourier transform of the quantum correlation
function.

\begin{defn}\label{VO.psd.def}
The power spectral density of a stationary environment operator vector $c$ is
defined as
\begin{equation}
  \label{xue.psd}
  S(\omega)= \int_{-\infty}^{+\infty} \langle c(t+\tau)c^\dagger(t)\rangle
  e^{-s t}dt.
\end{equation}
\end{defn}
Note that since the definition of Fourier transform for quantum systems is actually the standard inverse Fourier transform but double sided Laplace transform in equation (\ref{xue.psd}) is standard, we have $s=-{\rm i}\omega$.

It is straightforward to verify using singular commutation relations that
when $c$ is the quantum white noise $b(t)$ and the environment is
Markovian, then $S(\omega)=1$. On the other hand, when
$S(\omega)$ defined by equation (\ref{xue.psd}) is not equal to  1, this
corresponds to a colored noise $c$.

To model internal modes of the environment, we introduce an ancillary
Markovian quantum system with a Hamiltonian operator $\mathsf{H}_a$ and
a collection of coupling operators with respect to ancillary quantum
white noises, combined into a coupling operator vector $L_a$.
Using the compact $(S,L,\mathsf{H})$ notation, such an ancillary system is
denoted
\begin{equation}\label{7}
  \textsf{G}_a=({\rm I},L_a,\mathsf{H}_a);
\end{equation}
as noted previously, the scattering matrix is assumed to be identity.
Note that the ancillary system evolves on a Hilbert space
$\mathfrak{h}_a\otimes\mathfrak{F}_a$, where the ancillary system
and the corresponding white noise are defined on the Hilbert space
$\mathfrak{h}_a$ and the Fock space $\mathfrak{F}_a$, respectively.
Also associated with this system is a vector $c$ of operators of the
ancillary system defined on the Hilbert space $\mathfrak{h}_a$. This vector
of operators represents the quantum colored noise whereby the ancillary
system interacts with the principal system; see (\ref{9}). As the ancillary
system is driven by a quantum white noise, it can be thought of whitening
the quantum colored noise arising from the non-Markovian environment.

\begin{figure}
  \includegraphics[width=8.5cm]{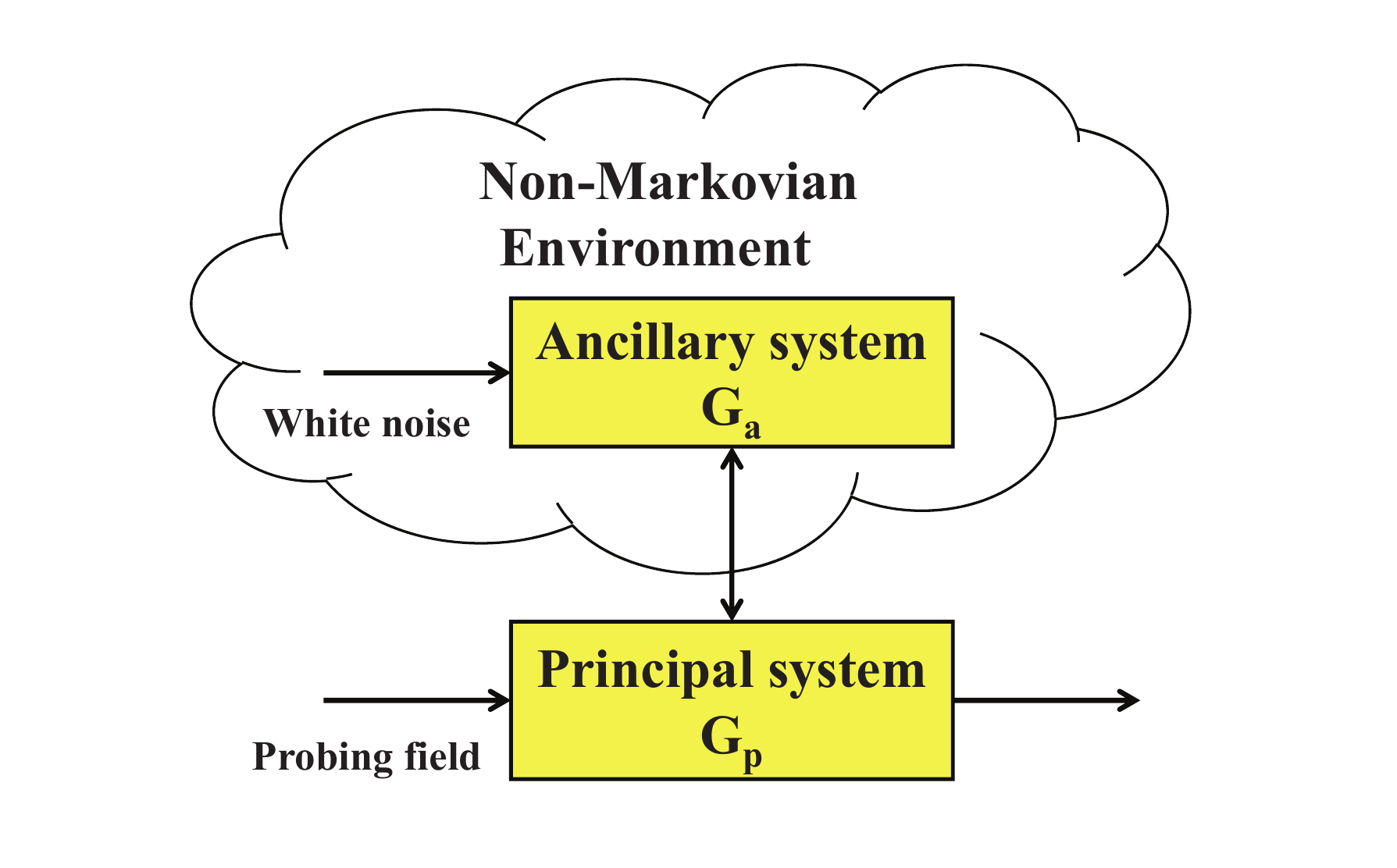}\\
  \caption{A schematic plot of the general augmented system model for non-Markovian quantum systems.}\label{AP2}
\end{figure}

The general quantum feedback network theory~\cite{Gough} can now be applied
to describe interactions between the principal system and the
environment. From this theory, the combined system model representing
the principal quantum system $\textsf{G}_p$ and the ancillary system model
(\ref{7}) for the non-Markovian environment interacting via the Hamiltonian
(\ref{9}) has the following $(S,L,\mathsf{H})$ parameters,
\begin{equation}\label{10}
  \textsf{G}_{T}=({\rm I},\left(
               \begin{array}{c}
                 L_a \\
                 L_p \\
               \end{array}
             \right), \mathsf{H}_p+\mathsf{H}_a+\mathsf{H}_{pa}).
\end{equation}
A schematic diagram of the augmented system is shown in Fig.~\ref{AP2}. The
augmented system $\textsf{G}_{T}$ is defined on the tensor product Hilbert space
$\mathfrak{h}_p\otimes\mathfrak{h}_a\otimes\mathfrak{F}_p\otimes\mathfrak{F}_a$. Since
this augmented system only interacts with quantum white noise fields,
namely with the ancillary white noise field and the probing field, the
overall system is
Markovian. However, as we will show in the next section, the principal
subsystem is not Markovian due to the interaction with the ancillary system.

\subsubsection{Quantum stochastic differential equations for the augmented
  system}

By substituting (\ref{10}) into (\ref{2}) the Heisenberg picture stochastic
differential equation of the evolution operator $U(t)$ for the augmented
system (\ref{10}) can be obtained to be
\begin{eqnarray}\label{11}
           {\rm d}U(t)&=&{\Big (}-{\rm i}(\mathsf{H}_p+\mathsf{H}_a+\mathsf{H}_{pa}){\rm d}t-\frac{1}{2}L_{p}^\dagger L_{p}{\rm d}t\nonumber\\
        && -\frac{1}{2}L_{a}^\dagger L_{a}{\rm d}t+{\rm d}B_p^\dagger(t) L_{p}-L_{p}^\dagger{\rm d}B_p(t)\nonumber\\
        &&+{\rm d}B_a^\dagger(t) L_{a}-L_{a}^\dagger{\rm d}B_a(t){\Big )}U(t).
\end{eqnarray}
Here, ${\rm d}B_p$ and ${\rm d}B_a$ are the quantum infinitesimal
increments for the noise processes of the principal and ancillary system,
respectively. Then using (\ref{11}), the evolution of an augmented system
operator $X'$ can be defined as $X'(t)=U^\dagger(t)X'U(t)$ which satisfies
the quantum stochastic differential equation
\begin{eqnarray}\label{12}
  {\rm d}X'(t) &=& -{\rm i}[X'(t),\mathsf{H}'(t)]{\rm d}t\nonumber\\
  &+& ( \mathcal{L}_{L_p(t)}(X'(t))+\mathcal{L}_{L_a(t)}(X'(t))){\rm d}t\nonumber\\
  &+&([X'(t),c^\dagger(t)z(t)]+[z^\dagger(t)c(t),X'(t)]){\rm d}t\nonumber\\
  &+&{\rm d}B_p^\dagger(t)[X'(t),L_{p}(t)]+[L_{p}^\dagger(t),X'(t)]{\rm d}B_p(t)\nonumber\\
  &+&{\rm d}B_a^\dagger(t)[X'(t),L_{a}(t)]+[L_{a}^\dagger(t),X'(t)]{\rm d}B_a(t),\nonumber\\
\end{eqnarray}
where $\mathsf{H}'(t)=U^\dagger(t)(\mathsf{H}_p+\mathsf{H}_a)U(t)$ and
other time-varying operators are obtained in the same way as $X'(t)$.

Note that any operator of the augmented system $X'$ can be written as
$X'=X_p\otimes X_a$, i.e., a tensor product of a principal system operator
$X_p$ and an ancillary system operator $X_a$. Thus, the generator for the
augmented system can be expressed as
\begin{equation}\label{13}
  \mathcal{G}_{T}=\mathcal{G}_p(X_p)\otimes X_a+X_p\otimes\mathcal{G}_a(X_a)-{\rm i}[X',\mathsf{H}_{pa}],
\end{equation}
where\begin{eqnarray}
       \mathcal{G}_p(X_p) &=& -{\rm i}[X_p,\mathsf{H}_p]+\mathcal{L}_{L_p}(X_p), \\
        \mathcal{G}_a(X_a) &=& -{\rm i}[X_a,\mathsf{H}_a]+\mathcal{L}_{L_a}(X_a)\label{gxa}
     \end{eqnarray}
     are the generators for the principal system and the ancillary system, respectively.

     In particular, for $X'=X_p\otimes {\rm I}$, i.e., a principal system operator, Eq.~(\ref{12})  reduces to
     \begin{eqnarray}\label{14}
        {\rm d}X_p(t) &=& -{\rm i}[X_p(t),\mathsf{H}_p(t)]{\rm d}t+ \mathcal{L}_{L_p(t)}(X_p(t))\nonumber\\
  &&+{\big (}c^\dagger(t)[X_p(t),z(t)]+[z^\dagger(t),X_p(t)]c(t){\big )}{\rm d}t\nonumber\\
  &&+{\rm d}B_p^\dagger(t)[X_p(t),L_{p}(t)]+[L_{p}^\dagger(t),X_p(t)]{\rm d}B_p(t)\nonumber\\
     \end{eqnarray}
     with $\mathsf{H}_p(t)=U^\dagger(t)\mathsf{H}_pU(t)$. When $X'={\rm I}\otimes X_a$, i.e., an operator of the ancillary system, we have
     \begin{eqnarray}\label{15}
       {\rm d}X_a(t) &=& -{\rm i}[X_a(t),\mathsf{H}_a(t)]{\rm d}t+\mathcal{L}_{L_a(t)}(X_a(t))){\rm d}t\nonumber\\
  &&+([X_a(t),c^\dagger(t)]z(t)+z^\dagger(t)[c(t),X_a(t)]){\rm d}t\nonumber\\
  &&+{\rm d}B_a^\dagger(t)[X_a(t),L_{a}(t)]+[L_{a}^\dagger(t),X_a(t)]{\rm d}B_a(t)\nonumber\\
     \end{eqnarray}
with $\mathsf{H}_a(t)=U^\dagger(t)\mathsf{H}_aU(t)$. Note that in quantum mechanics it is conventional to write $X_p\otimes {\rm I}$ and ${\rm I}\otimes X_a$ as $X_p$ and  $X_a$, respectively.

One can see form~(\ref{14}) that due to the direct coupling term induced by
the ancillary system in the second line of Eq.~(\ref{14}), the principal
system will not behave as a Markovian quantum system when it is coupled
with the ancillary system. Indeed, the ancillary
system operator $c$ can be treated as an input into the principal
system generated by the ancillary system. As we noted earlier this input
represents a colored noise generated by the environment, which explains a
non-Markovian nature of the principal systems dynamics. Also, it can be seen
from~(\ref{15}) that the evolution of the ancillary system operator $X_a$
depends on the principal system operator $z$, which shows that the principal
system acts back on the ancillary system. This explains the non-Markovian
behaviour of the environment.

\subsubsection{A master equation for the augmented system}
Alternatively, dynamics of the augmented system can be described using a
master equation. In particular, it is convenient to use master equations
when the principal system is a qubit system involving nonlinear
dynamics. This will be shown in section~\ref{sec5}.

For the augmented system (\ref{10}), the master equation (\ref{5-3}) takes
the form
\begin{eqnarray}\label{16}
 \dot \rho(t)&=&-{\rm i}[\mathsf{H}_p+\mathsf{H}_a,\rho(t)]+\mathcal{L}^*_{L_a}(\rho(t))+\mathcal{L}^*_{L_p}(\rho(t))\nonumber\\
 &&+[c^\dagger z,\rho(t)]+[\rho(t), z^\dagger c],
\end{eqnarray}
where $\rho(t)$ is the density matrix of the augmented system.
Once again we observe that the state evolution of the augmented system is
Markovian, since future values of the density matrix $\rho(t)$ only depend
on the present density matrix. One can also obtain the density matrix
$\rho_p(t)$ of the principal system as the partial trace with
respect to the ancillary system,
\begin{equation}\label{17}
  \rho_p(t)={\rm tr}_a[\rho(t)].
\end{equation}

\subsection{Linear quantum systems models for non-Markovian
  environments with rational power
  spectral densities}\label{subs32}

We now demonstrate that the proposed modelling of non-Markovain
systems is sufficiently rich in a sense that for a broad class of
environment power spectral densities $S(\omega)$, an ancillary system can
be constructed whose characteristics match $S(\omega)$. Specifically, we
show that when the environment has a rational power spectral
density, the corresponding ancillary system can be realized within the class of
linear quantum systems. Clearly, when $S(\omega)$ is not rational, but can
be approximated by a rational power spectral density, an approximating
ancillary linear quantum system model can be constructed. From a practical
perspective, such an approximation is often sufficient and leads to
meaningful results as will be demonstrated in Section~\ref{sec5}.


First, let us consider a linear quantum system comprised of $n$ harmonic
oscillators interacting with $m$ channels of quantum white noise
fields and obtain an expression of the associated power spectral
density. Since our aim is to represent a non-Markovian environment in a
form of an ancillary system and such environments do not generate energy, we
restrict attention to the class of linear quantum systems whose
quantum stochastic differential equations only involve annihilation
operators.

To obtain an expression for the power spectral density of a linear
annihilation only system, we begin with its $(S,L,\mathsf{H})$
description. Specifically, the Hamiltonian of such a linear system has the
form
\begin{equation}\label{23}
\mathsf{H}_a=a^\dagger\Omega a,
\end{equation}
where $a=[a_1,a_2,\cdots,a_n]^T$ is a column vector of annihilation
operators with an annihilation operator $a_j$ as its $j$-th
component, and $a^\dagger=[a_1^\dagger,a_2^\dagger,\cdots,a_n^\dagger]$ is the
corresponding row vector of creation operators. These operators are defined
on the common Hilbert space $\mathfrak{h}_a$ and satisfy the
singular commutation relations
\begin{equation}\label{24}
  [a_j,a_k^\dagger]=\delta_{jk},~~[a_j,a_k]=0,~~j,k=1,\cdots,n.
\end{equation}
The diagonal and non-diagonal elements of the Hermitian matrix
$\Omega\in\mathbb{C}^{n\times n}$ represent the internal angular
frequencies and the couplings between the harmonic oscillators,
respectively. Also, the system is coupled with a white noise field, and
the corresponding coupling operator $L_a$ of such a linear annihilation
only system  with respect to the quantum white noise fields $b_a(t)$ can be
expressed as
\begin{equation}\label{24-1}
  L_a=N_aa
\end{equation}
with a matrix $N_a\in\mathbb{C}^{m\times n}$. Also, according to our standing assumption, the
identity scattering matrix is assumed.

With these Hamiltonian and coupling operators, the evolution operator
$\bar{U}(t)$ of  the linear ancillary system satisfies
\begin{eqnarray}\label{25}
{\rm d}{\bar{U}}(t)&=&\left(-\left(\frac{1}{2}L_a^\dagger L_a+{\rm
      i}a^\dagger\Omega a\right){\rm d}t\right. \nonumber \\
&&\left.\phantom{\frac{1}{2}}+{\rm d}B_a^\dagger(t)L_a-L_a^\dagger {\rm d}B_a(t)\right)\bar U(t),
\end{eqnarray}
where ${\rm d}B_a$ is the quantum infinitesimal
increment for the white noise field process. Then, the annihilation operators
$a(t)=\bar{U}^\dagger(t) a\bar{U}(t)$
of the system evolve according to the linear quantum stochastic
differential equation
\begin{equation}\label{26}
{\rm d}a(t) =F_aa(t){\rm d}t+G_a{\rm d}B_a(t),
\end{equation}
where $F_a= -{\rm i}\Omega-\frac{1}{2}N_a^\dagger N_a$ and $G_a=-N_a^\dagger$.

Our result concerning the modelling of a colored noise environment in
terms of linear systems of the form (\ref{26}) is summarized in the following
theorem. Since non-Markovian quantum systems normally involve only one kind
of colored noise, for simplicity we restrict attention to single input ancillary
systems;  that is, $m=1$ and the corresponding spectral density $S(\omega)$ is
scalar. The extension to the matrix case is quite trivial, as will be seen
from the proof.

\begin{thm}\label{VO.T5}
Suppose that the power spectral density $S(\omega)$ of an environment
colored noise process is rational and satisfies $S(\omega)\geq 0$
for all $\omega$. Then there exists a linear quantum system in the form of
(\ref{26}) and a matrix $H_a\in\mathbb{C}^{1 \times n}$, such
that the process
\begin{equation}\label{27}
  c(t)=H_a a(t)
\end{equation}
has the desired power spectral density $S(\omega)$.
\end{thm}

Note that although the operator (\ref{27}) is expressed in terms of
operators of the system (\ref{26}), unlike the input defined in
Eq.~(\ref{5}) it is not an output field of the system
available for quantum measurement. Yet it represents a physical
quantity whereby the ancillary system can interact with other quantum
systems.

The proof of Theorem~\ref{VO.T5} will be given later in this section.
Before proving the theorem, it is instructive to compute the power spectral
density of the operator (\ref{27}).
Since the system (\ref{26}) is to represent internal modes of the environment,
its dynamics can be assumed to start from a long time ago. Formally, this
means that the initial time $t_0$ when the system (\ref{26}), (\ref{27})
was at rest is $-\infty$. Hence, $c(t)$ can be expressed as
\begin{equation}\label{29}
  c(t)=\int_{-\infty}^t\Xi_a(t-\theta){\rm d}B_a(\theta),
\end{equation}
where
\begin{equation}\label{29-1}
\Xi_a(t)=\begin{cases}H_a e^{-F_a t}G_a, & t\ge 0; \\
                      0, & t< 0
                    \end{cases}
\end{equation}
is the inverse Laplace transform of the $r\times m$ transfer function matrix
\begin{equation}\label{30}
\Gamma(s)=H_a(s{\rm I}-F_a)^{-1}G_a.
\end{equation}
That is, $c(t)$ is analogous to the stationary response of a
linear system with the transfer function matrix $\Gamma(s)$ to a
white noise input. 

The covariance of $c(t)$ can be found using the quantum Ito calculus
rules. Assuming that $\tau\ge 0$ (the case $\tau< 0$ is treated in the same
way),  we obtain
\begin{eqnarray*}
\lefteqn{\langle c(t+\tau)c^\dagger (t)\rangle} && \\
&=&
\left\langle \int_{-\infty}^{t+\tau} \Xi_a(t+\tau-\theta) {\rm d}B_a(\theta)
\int_{-\infty}^t{\rm d}B_a^\dagger (\theta) \Xi_a^\dagger
(t-\theta)\right\rangle \\
&=&
\left\langle \int_{-\infty}^t \Xi_a(t+\tau-\theta) {\rm d}B_a(\theta)
\int_{-\infty}^t{\rm d}B_a^\dagger (\theta) \Xi_a^\dagger (t-\theta) \right\rangle\\
&&
+ \left\langle \int_t^{t+\tau} \Xi_a(t+\tau-\theta) {\rm d}B_a(\theta)
\int_{-\infty}^t{\rm d}B_a^\dagger (\theta) \Xi_a^\dagger (t-\theta)\right\rangle
\\
\end{eqnarray*}
The last integral vanishes since the time intervals do not overlap and thus we have
\begin{eqnarray*}
\lefteqn{\langle c(t+\tau)c^\dagger (t)\rangle} && \\
&=&
\int_{-\infty}^t \Xi_a(t+\tau-\theta)\Xi_a^\dagger (t-\theta) d\theta
\\
&=&
\int_{-\infty}^t \Xi_a(t+\tau-\theta)\Xi_a^\dagger (t-\theta) d\theta \\
&=&
\int_0^{\infty} \Xi_a(\tau+\lambda)\Xi_a^\dagger (\lambda) d\lambda \\
&=&
\int_{-\infty}^{\infty} \Xi_a(\tau+\lambda)\Xi_a^\dagger (\lambda) d\lambda
\end{eqnarray*}
Using Definition~\ref{VO.psd.def}, its power spectral density
can be computed to be
\begin{eqnarray}\label{31}
\lefteqn{
\int_{-\infty}^{+\infty}\langle c(t+\tau)c^\dagger (t)\rangle
  e^{-s \tau}d\tau} && \nonumber\\
&=&
\int_{-\infty}^{+\infty}\int_{-\infty}^{+\infty} \Xi_a(\tau+\lambda)\Xi_a^\dagger
(\lambda)e^{-s \tau} d\lambda d\tau \nonumber\\
&=&\int_{-\infty}^{+\infty}\int_{-\infty}^{+\infty} \Xi_a(\tau+\lambda)
e^{-s (\tau+\lambda)} \Xi_a^\dagger
(\lambda)e^{s \lambda} d\lambda d\tau \nonumber\\
&=&
\int_{-\infty}^{+\infty} \Xi_a(\tau+\lambda)
e^{-s (\tau+\lambda)} d\tau  \int_{-\infty}^{+\infty}
\Xi_a^\dagger (\lambda)e^{s \lambda} d\lambda \nonumber\\
&=&\int_{-\infty}^{+\infty} \Xi_a(\tau+\lambda)
e^{-s (\tau+\lambda)} d\tau \left( \int_{-\infty}^{+\infty}
\Xi_a (\lambda)e^{-(-s^*) \lambda} d\lambda\right)^\dagger\nonumber\\
&=&\Gamma(s)\Gamma^\sim (s).
\end{eqnarray}
where $\Gamma^\sim(s)=\Gamma^\dagger(-s^*)$ is the adjoint of the
transfer function $\Gamma(s)$. The calculation is analogous to the
calculation of the power spectral density
for an output of a classical linear system driven by a white noise
input~\cite{KS72}.

It follows from (\ref{31}) that the spectral factorization method can be
employed to obtain a linear quantum system representation of the
environment with a positive rational spectral density $S(\omega)$.

\paragraph*{Proof of Theorem~\ref{VO.T5}}
Given a power spectral density $S(\omega)$, we will determine the
corresponding Hamiltonian (\ref{23}) and the coupling operator (\ref{24-1})
which define the linear quantum system (\ref{26}), (\ref{27}). First we
observe that according to \cite[Theorems 5, 7]{JCW}, $S(\omega)$ can be
factorized as in (\ref{31}) with a stable transfer function
\begin{equation}\label{32}
   \Gamma(s)=\frac{\beta_{n-1} s^{n-1}+\cdots+\beta_0}{s^n+\alpha_{n-1}s^{n-1}+\cdots+\alpha_0}.
\end{equation}
Next, a stable transfer function $\Gamma(s)$ of the form (\ref{32}) has a
state-space realization with matrices
\begin{eqnarray}\label{33}
  F_0 &=& \left[
            \begin{array}{cccc}
              0 & 1 & \cdots &0 \\
               & \ddots & \ddots &  \\
              0 &  &  & 1\\
              -\alpha_0 & -\alpha_1 & \cdots & -\alpha_{n-1} \\
            \end{array}
          \right]
  ,~~G_0=\left[
           \begin{array}{c}
             0 \\
             \vdots \\
             0 \\
             1 \\
           \end{array}
         \right],
  \nonumber \\
  H_0 &=& \left[
            \begin{array}{ccc}
              -\beta_0 ,& \cdots ,& -\beta_{n-1} \\
            \end{array}
          \right],
\end{eqnarray}
i.e., $\Gamma(s)=H_0(s{\rm I}-F_0)^{-1}G_0$; e.g., see~\cite{KS72}. Since
such a realization is Hurwitz and controllable, the Lyapunov equation
\begin{equation}\label{34-1}
F_0 P+ P F_0^\dagger+G_0 G_0^\dagger=0
\end{equation}
for this realization has a unique and invertible solution
$P>0$~\cite{KS72}. Hence, we can find a factorization for the inverse of
$P$, i.e., $P^{-1}=T^\dagger T$ and thus equation (\ref{34-1}) can be
reexpressed  as
\begin{equation}\label{34-1-1}
  TF_0T^{-1}+(T^{-1})^\dagger F_0^\dagger T^\dagger+TG_0 G_0^\dagger T^{\dagger}=0.
\end{equation}
Define
\begin{equation}\label{33-1}
  F_a=TF_0T^{-1},~~G_a=TG_0,~~H_a=H_0T^{-1}.
\end{equation}
and substituting this notation into (\ref{34-1-1}):
\begin{equation}\label{33-2}
  F_a+ F_a^\dagger+G_aG_a^\dagger=0,
\end{equation}
This equation shows the new realization (\ref{33-1}) obtained from
(\ref{33}) by applying the coordinate transformation $T$ satisfies the
physical realizability condition of~\cite{maalouf}.
Also, it follows from~\cite{maalouf} that we can obtain expressions for the
Hermitian matrix $\Omega$ and the coupling operator $L_a$ as
\begin{equation}\label{33-5-1}
  \Omega=\frac{\rm i}{2}(F_a-F_a^\dagger), \quad   L_a=-G_a^\dagger a.
\end{equation}
These quantities define the system (\ref{26}) as a linear quantum
system. \hfill $ \Box $

As an example, let us particularize the result of Theorem~\ref{VO.T5} for the
Lorentzian power spectral density of the form
\begin{equation}\label{36-1}
  S_0(\omega)=\frac{\frac{\gamma_0^2}{4}}{\frac{\gamma_0^2}{4}+(\omega-\omega_0)^2},
\end{equation}
which commonly arises in solid-state systems~\cite{WeiMinPRL2012,Tu2008}.

\begin{coro}\label{coro}
The power spectral density (\ref{36-1}) of the Lorentzian noise can be realized
by a single mode linear quantum system
\begin{eqnarray}
\mathrm{d}
a_0(t)&=&-(\frac{\gamma_0}{2}+{\rm
  i}\omega_0)a_0(t)dt-\sqrt{\gamma_0}\mathrm{d}B_a(t), \label{40-1-1} \\
 c_0(t)&=&-\frac{\sqrt{\gamma_0}}{2}a_0(t), \nonumber
\end{eqnarray}
with the annihilation operator $a_0$.
\end{coro}

\begin{pf}
The Lorentzian spectrum (\ref{36-1}) can be factorized as in (\ref{31}) with
\begin{equation}\label{37}
  \Gamma_0(s)=\frac{\frac{\gamma_0}{2}}{s+{\rm i}\omega_0+\frac{\gamma_0}{2}}.
\end{equation}
The order of the denominator of this transfer function is one, i.e.,
$n=1$, which means the ancillary system can be realized by a single mode
linear quantum system. Then, the matrices $F_0$, $G_0$, $H_0$ and $T$
reduce to scalars. From (\ref{33}), we obtain
\begin{equation}\label{38}
  F_0=-(\frac{\gamma_0}{2}+{\rm i}\omega_0),~~G_0=1,~~H_0=\frac{\gamma_0}{2}.
\end{equation}
Substituting (\ref{38}) into the Lyapunov equation (\ref{34-1-1}), we have $T^\dagger T=\gamma_0$, which can be solved to give
\begin{equation}\label{39}
T=-\sqrt{\gamma_0}.
\end{equation}
According to (\ref{33-1}), the realization of
the system (\ref{26}) with
\begin{equation}\label{40}
  F_a=-(\frac{\gamma_0}{2}+{\rm i}\omega_0),~~G_a=-\sqrt{\gamma_0},~~H_a=-\frac{\sqrt{\gamma_0}}{2},
\end{equation}
satisfies the physical realizability condition (\ref{33-2}). Hence,
equation (\ref{40-1-1}) corresponds to a
physically realizable linear quantum system. Indeed, from (\ref{33-5-1}),
the Hamiltonian and the coupling operator of this
system can be obtained
\begin{equation}\label{41}
  \mathsf{H}_a=\omega_0a ^\dagger_0 a_0, \quad L_a=\sqrt{\gamma_0}a_0
\end{equation}
where $a_0$ is the annihilation operator of the system.\hfill $ \Box $
\end{pf}

\begin{rem}
The central frequency of the Lorentzian power spectral density (\ref{36-1})
determines the angular frequency $\omega_0$ of the ancillary system,  and
the bandwidth of the power spectral density determines the system damping rate
$\gamma_0$ with respect to the white noise field.
\end{rem}

\section{Application to quantum filtering}~\label{sec4}

\subsection{Whitening quantum filter for non-Markovian quantum systems}

In quantum physics, to force a quantum system to generate an
output field, it must be excited with a probing field. Typically, such a
filed is a quantum white noise field, $b_p(t)$.
The quadrature of the output field,
\begin{equation}\label{43-7-1}
  Y(t)=U^\dagger(t)(B_p+B^\dagger_p)U(t),
\end{equation}
can be monitored via homodyne detection, and the measurement results can
be utilized to construct a filter for the system. In this section we show
that when the probing field is
applied to a principal system directly interacting with an ancillary system,
an augmented-system-based quantum filter, i.e., a
whitening quantum filter can be derived to process measurements of the
output field. Our
derivation is based on the following assumptions regarding the physical
apparatus.

\begin{assumption}\label{assm1}
\begin{enumerate}[1.]
\item
The probing field is a quantum white noise field in a vacuum state, which
satisfies a non-demolition condition such that the dynamics of the
augmented system can be continuously monitored~\cite{bouten}.

\item
The monitored channels are assumed to be coupled with the principal system
via the operator $L_p$.

\item
The homodyne detector is perfect with $100\%$ detection efficiency.
\end{enumerate}
\end{assumption}

\begin{defn}\label{VO.QF.def}
The quantum filtering problem is to determine an estimate of an observable
$ X'(t)$ which is the conditional expectation
\[
\hat X'(t)=\pi_t( X')=\mathbb{E}[ X'(t)|\mathcal{Y}(t)],
\]
i.e, the projection of $X'(t)$ on
$\mathcal{Y}(t)$, the commutative subspace of operators generated by the
measurement results $Y(\tau),~0\leq\tau\leq t$.
\end{defn}

By applying the existing quantum filtering theory, we immediately obtain
the whitening filter for the augmented system derived in Section~\ref{sec3}.

\begin{thm}
Under the Assumption \ref{assm1}, a quantum filter for the
augmented system can be constructed as
\begin{eqnarray}\label{43-7-2}
    {\rm d}\pi_t( X')&=& \pi_t(\mathcal{G}_T (X')){\rm d}t-(\pi_t( X' L_p+{ L^\dagger_p}  X')-\pi_t( X')\nonumber\\
    &&\times\pi_t( L_p+ {L_p^\dagger}))({\rm d}Y(t)-\pi_t( L_p+
    {L_p^\dagger})  {\rm d}t).
\end{eqnarray}
The measurement process $Y(t)$ induces an innovation process $W(t)$
satisfying
\[
{\rm d}W(t)={\rm d}Y(t)-\pi_t( L_p+ {L_p}^\dagger)  {\rm d}t
\]
 and whose increment ${\rm d}W(t)$ is independent of $\pi_\tau( X')$,
$\tau\in[0, t]$.
\end{thm}

\begin{pf}
This whitening filter can be derived by applying the standard orthogonal
projection approach developed in~\cite{bouten,Belavkin} to the augmented
system model $\mathsf{G}_T$. \hfill $ \Box $
\end{pf}

Our next result concerns estimating the density matrix of the augmented
system.

\begin{defn}\label{defcdm}
The conditional density matrix $\hat \rho(t)$ of a quantum system is
a density matrix which satisfies the equation
\begin{equation}\label{VO.cond.density}
\pi_t( X')={\rm tr}[\hat \rho(t) X'].
\end{equation}
\end{defn}

\begin{rem}
According to the above definition, the conditional density matrix is a
density matrix for which the quantum expectation of $X'$ in the Schr{\"o}dinger
picture coincides with the orthogonal projection of $X'(t)$ onto the
commutative subspace $\mathcal{Y}(t)$.
\end{rem}

\begin{thm}
The conditional density matrix $\hat \rho(t)$ for the augmented system satisfies
the stochastic master equation
\begin{equation}\label{43-7-3}
 {\rm d}\hat \rho(t) =\mathcal{G}_T^*(\hat \rho(t)){\rm d}t+\mathcal{F}_{ L_p}(\hat \rho(t)){\rm d}W
\end{equation}
with
\begin{equation}\label{43-7-4}
\mathcal{F}_{ L_p}(\hat \rho(t))= L_p\hat \rho(t)+\hat \rho(t) {L_p^\dagger}-{\rm tr}[( L_p+ {L_p^\dagger})\hat \rho(t)]\hat \rho(t),
\end{equation}
where the superoperator $\mathcal{G}_T^*={\rm i}[H_a+H_p+H_{pa},\hat \rho(t)]+\mathcal{L}_{L_p}^*(\hat \rho(t))+\mathcal{L}_{L_a}^*(\hat \rho(t))$ is the adjoint of
$\mathcal{G}_T$.
\end{thm}

%
\begin{pf}
The stochastic master equation for the conditional density matrix defined
in (\ref{VO.cond.density}) can be derived from the whitening quantum filter
(\ref{43-7-2}) by using the methods in~\cite{bouten,Belavkin}. \hfill $\Box $
\end{pf}

In practice, when one is interested in estimating the principal
non-Markovian system, it is the conditional density matrix of the principal
system $\hat \rho_p(t)$ that will be of interest. Formally, it can be obtained
as the partial trace of $\hat \rho(t)$, by tracing out the ancillary system
\begin{equation}\label{43-7-5}
  \hat \rho_p(t)={\rm tr}_a[\hat \rho(t)].
\end{equation}
Earlier in Section~\ref{subs32} we have proposed representing ancillary
systems using linear quantum systems. In this case, equation (\ref{43-7-5})
reduces to $\hat
\rho_p(t)=\sum_n\langle n|\hat\rho(t)|n\rangle$ where the number of bases
$|n\rangle$ for the linear quantum system is infinite. This makes obtaining
an exact expression for $\hat \rho_p(t)$ from (\ref{43-7-5}) rather difficult.
However, approximating the linear ancillary system with a $N$-level
system will have an effect of truncation, and thus it is possible to
calculate an approximation to the partial trace
(\ref{43-7-5})~\cite{6189045}.

\subsection{Whitening quantum filter for linear non-Markovian quantum
  systems}

In this section, we consider the case where the principal system is a
linear quantum system whose description involves only annihilation
operators. For such quantum systems, we show that the
whitening quantum filter is actually a quantum Kalman filter.


As noted previously, the Hamiltonian of a linear annihilation only
principal system is quadratic and can be expressed as
\begin{equation}\label{47}
  \mathsf{H}_p=d^\dagger\Lambda d,
\end{equation}
where $d=[d_1,d_2,\cdots,d_{n'}]^T$ is a column vector of annihilation
operators for the principal system. Also as before,
$d^\dagger=[d_1^\dagger,d_2^\dagger,\cdots,d_{n'}^\dagger]$ is the
corresponding row vector of creation operators for the principal system.
These operators satisfy the singular commutation relations, cf.~(\ref{24}):
\begin{equation}\label{48}
  [d_j,d_k^\dagger]=\delta_{jk},~~[d_j,d_k]=0,~j,k=1,\cdots,n'.
\end{equation}
The diagonal and off-diagonal elements of the Hermitian matrix $\Lambda\in\mathbb{C}^{n'\times n'}$ are determined by the angular frequencies of components of the principal system and their couplings, respectively.

Also, the coupling operator with respect to $m'$ channels of the quantum white noise process ${\rm d}B_p(t)$ 
 and the direct coupling operator are specified as
\begin{equation}\label{49}
L_p=N_pd,~~ z=K_pd,
\end{equation}
respectively, where $N_p\in\mathbb{C}^{m'\times n'}$ and $K_p\in\mathbb{C}^{r\times n'}$.

As described in the previous section, the system interacts with a
non-Markovian environment. These interactions can be captured using a model
where the principal system is directly coupled with a linear ancillary
system. Thus, the quantum stochastic differential equation
 for the augmented system including both the principal and the ancillary
 systems can be obtained
\begin{eqnarray}\label{50}
  \left[
    \begin{array}{c}
      {\rm d} \dot d(t) \\
      {\rm d}\dot{a}(t) \\
    \end{array}
  \right]&=&\left[
            \begin{array}{cc}
              F_p & -K_p^\dagger H_a \\
              H_a^\dagger K_p & F_a \\
            \end{array}
          \right]\left[
                   \begin{array}{c}
                     d(t){\rm d}t \\
                     {a}(t) {\rm d}t \\
                   \end{array}
                 \right]\nonumber\\
                 &&~~~~~~~~~~~~~~~~+\left[
                           \begin{array}{cc}
                             G_p &0\\
                             0 & G_a \\
                           \end{array}
                         \right]\left[
                                  \begin{array}{c}
                                  {\rm d}B_p(t) \\
                                   {\rm d}B_a(t) \\
                                  \end{array}
                                \right],
\end{eqnarray}
where $F_p= -{\rm i}\Lambda-\frac{1}{2}N_p^\dagger N_p$ and
$G_p=-N_p^\dagger$. Also, the output field excited by the probing the
principal system with the white noise filed process ${\rm d}B_p(t)$ is
\begin{equation}\label{51}
  {\rm d}B_{{\rm out}-p}(t)=H_pd(t){\rm d}t+{\rm d}B_p(t)
\end{equation}
with $H_p=N_p$.

Suppose that the position quadrature of the output,
\[
{\rm
  d}y(t)=\frac{1}{\sqrt{2}}( {\rm d}B_{{\rm out}-p}(t)+ {\rm d}B_{{\rm
    out}-p}^\dagger(t))
\]
 is observed via homodyne detection. Also, since the operators in
 Eqs.~(\ref{50}) and~(\ref{51}) are not self-adjoint and
the coefficients may be complex valued, it is convenient to transform
Eqs.~(\ref{50}) and~(\ref{51}) into a phase space in terms of position and
momentum operators as
\begin{eqnarray}
  {\rm d}{ x}(t)
   &=& A { x}(t){\rm d}t+{\rm d}w_1(t),\label{52-1}\\
 {\rm d}y(t)
   &=&C{ x}(t){\rm d}t+{\rm d}w_2(t)
\end{eqnarray}
with ${\rm d}w_1(t)=B {\rm d}w(t)$, ${\rm d}w_2(t)=D {\rm d}w(t)$,
\begin{eqnarray*}\label{53}
     A &=& \frac{1}{2}\left[
             \begin{array}{cccc}
A_{11} & A_{12} & A_{13} A_{14} \\
A_{21} & A_{22} & A_{23} A_{24} \\
A_{31} & A_{32} & A_{33} A_{34} \\
A_{41} & A_{42} & A_{43} A_{44} \\
             \end{array}
           \right],\nonumber\\
A_{11}&=&A_{22}=F_p+F_p^\dagger, \quad  A_{12}=-A_{21}= {\rm
  i}(F_p-F_p^\dagger) \\
A_{13}&=&-K_p^\dagger H_a-K_p^T H_a^\dagger, \quad A_{14}=-{\rm
  i}(K_p^\dagger H_a-K_p^T H_a^\dagger), \\
A_{23}&=&{\rm i}(K_p^\dagger H_a-K_p^T H_a^\dagger), \quad
A_{24}=-K_p^\dagger H_a-K_p^T H_a^\dagger, \\
A_{31}&=&H_a^\dagger K_p+H_a^T K_p^\dagger, \quad A_{32}={\rm
  i}(H_a^\dagger K_p-H_a^T K_p^\dagger), \\
A_{33}&=&A_{44}=F_a+F_a^\dagger, \quad  A_{34}=-A_{43}= {\rm
  i}(F_a-F_a^\dagger) \\
A_{41}&=& -{\rm i}(H_a^\dagger K_p-H_a^T K_p^\dagger), \quad A_{42}=H_a^\dagger K_p+H_a^T K_p^\dagger \\
     B &=& \frac{1}{2}\left[
             \begin{array}{cccc}
B_{11} & B_{12} & B_{13} B_{14} \\
B_{21} & B_{22} & B_{23} B_{24} \\
B_{31} & B_{32} & B_{33} B_{34} \\
B_{41} & B_{42} & B_{43} B_{44} \\
             \end{array}
           \right],\nonumber\\
B_{11}&=&B_{22}=G_p+G_p^\dagger, \quad  B_{12}=-B_{21}= {\rm
  i}(G_p-G_p^\dagger) \\
B_{33}&=&B_{44}=G_a+G_a^\dagger, \quad  B_{34}=-B_{43}= {\rm
  i}(G_a-G_a^\dagger) \\
B_{13}&=&B_{14}=B_{23}+B_{24}=B_{31}=B_{32}=B_{41}=B_{42}=0 \\
    C&=& \frac{1}{2}\left[
                      \begin{array}{cccc}
                       H_p+H_p^\dagger &0&0&0\\
                      \end{array}
                    \right],\nonumber\\
                    D&=&\left[
                      \begin{array}{cccc}
                       {\rm I} &0&0&0\\
                      \end{array}
                    \right]\nonumber
   \end{eqnarray*}
where
\begin{eqnarray*}
x(t)&=&[q_p(t),p_p(t),q_a(t),p_a(t)]^T, \\ 
{\rm d}w(t)&=&[{\rm d}v_q(t),{\rm d}v_p(t),{\rm d}\bar v_q(t),{\rm d}\bar v_p(t)]^T
\end{eqnarray*}
are the quadrature representations of the operators of the principal and ancillary systems and the probing and quantum white noise fields, respectively.
The components of $ x(t)$ and $w(t)$ are calculated by applying the
coordinate transformation matrix $\Pi=\frac{1}{\sqrt{2}}\left[
                          \begin{array}{cc}
                            \rm I & \rm I \\
                            -{\rm i}\rm I &{\rm i}\rm I \\
                          \end{array}
                        \right]$ to the corresponding components of the
                        vectors of annihilation operators of the principal
                        and ancillary systems:
\begin{eqnarray*}
                        &&[ q_p(t),p_p(t)]^T=\Pi[d_p(t),d_p^\dagger(t)]^T, \\
                        &&[q_a(t),p_a(t)]^T=\Pi[a(t),a^\dagger(t)]^T, \\
&&[{\rm d}v_q(t),{\rm d}v_p(t)]^T=\Pi[{\rm d}B_p(t),{\rm
  d}B_p^\dagger(t)]^T, \\
&&[{\rm d}\bar v_q(t),{\rm d}\bar v_p(t)]^T=\Pi[{\rm d}B_a(t),{\rm d}B^\dagger_a(t)]^T.
\end{eqnarray*}
Note that the noise processes $w_1(t)$ and $w_2(t)$ are correlated, and the
covariance matrix of $[w_1(t),~w_2(t)]^T$ is $Vt$, where
\begin{equation}\label{53-1}
  V=\left[
      \begin{array}{cc}
        V_1 & V_{12} \\
        V_{12}^T & V_2 \\
      \end{array}
    \right]=\left[
              \begin{array}{cc}
                BB^T & BD^T \\
                DB^T & DD^T \\
              \end{array}
            \right].
\end{equation}

We are now in the position to present the whitening quantum filter for
computing the estimate $\hat x(t)$ of the vector of dynamical variables $x(t)$
of the linear augmented system model  (\ref{52-1}).

\begin{thm}
The whitening quantum filter for the linear augmented system (\ref{52-1})
is a steady-state quantum Kalman filter
\begin{eqnarray}
  {\rm d}{\hat x}(t) &=& A\hat x(t){\rm d}t+K({\rm d}y(t)-C\hat x(t){\rm
    d}t),\label{54}
\end{eqnarray}
with the Kalman gain
\[
K=(\hat VC^T+V_{12})V_2^{-1},
\]
obtained from the solution of the algebraic Riccati equation
\begin{eqnarray}
  \lefteqn{( A-V_{12}V_2^{-1}C)\hat V+\hat V(A-V_{12}V_2^{-1}C)^T} &&
  \nonumber\\
  &&\qquad-\hat V C^TV_2^{-1}C\hat V+V_1-V_{12}V_2^{-1}V_{12}^T=0. \label{54-1}
\end{eqnarray}
\end{thm}

This theorem is obtained by applying the existing quantum Kalman
filtering results for linear quantum
systems~\cite{Kjacobs,PhysRevA.74.032107} to the  linear augmented
Markovian system (\ref{52-1}). The obtained whitening quantum filter is
driven by the output $y(t)$. Note that the symmetrized covariance matrix
$\hat V$ is  independent of the measurement process.


\subsection{An illustrative example: A non-Markovian optical system}

\begin{figure}
  \includegraphics[width=8.5cm]{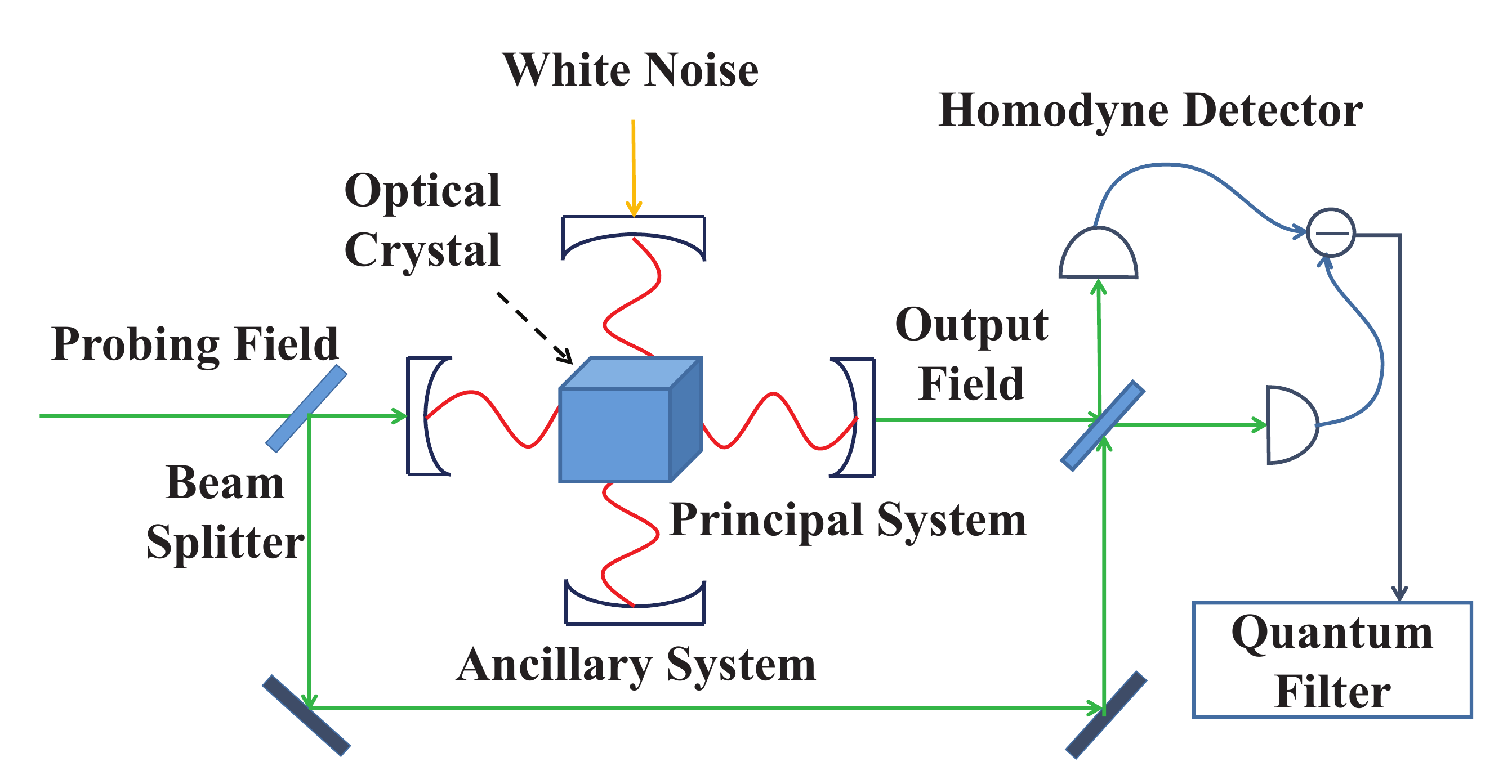}\\
  \caption{The illustrative example involving an optical system. The
    quantity of interest is the quantum state of the principal cavity. The
    ancillary cavity plays the role of the internal modes of the
    environment converting quantum white noise to Lorentzian noise and
    enabling estimation of the state of the principal cavity using the
    whitening quantum filter.}\label{example-opc}
\end{figure}

\paragraph*{The system}
In this example, we consider a single mode non-Markovian bosonic quantum
system disturbed by Lorentzian noise, which can be realized by two coupled
cavities as shown in Fig.~\ref{example-opc}. In this figure, the
horizontally oriented cavity is the principal system to be estimated.
The vertically oriented cavity is the ancillary system converting
white noise to Lorentzian noise. The optical modes in the two cavities are
directly and strongly coupled by an optical crystal. A probing field is
applied to the principal cavity, whose output is observed via homodyne
detection. These measurements are used as the data for a quantum filter.

The Hamiltonian operators of the principal and ancillary cavities are
$\mathsf{H}_p=\omega_p d^\dagger_p d_p$ and $\mathsf{H}_a=\omega_a
a^\dagger_0 a_0$, respectively, with angular frequencies
$\omega_p$ and $\omega_a$. They are expressed in terms of respective
annihilation and creation operators $d_p$, $d_P^\dagger$ and $a_0$,
$a_0^\dagger$ satisfying the corresponding singular commutation relations
(\ref{48}) and (\ref{24}). Then, the coupling operators of the principal
cavity $z$ and $L$ are specified to be $z=\sqrt{\kappa}d_p$ and
$L_p=\sqrt{\gamma_1}d_p$, with constant $\kappa$,
$\gamma_1$. Also, the coupling operator of the ancillary system with
respect to the white noise is $L_a=\sqrt{\gamma_0}a_0$. The cavities
interact through the operators $z$ and $c=-\frac{\sqrt{\gamma_0}}{2}a_0$;
see (\ref{9}). With these definitions, the quantum stochastic differential
equations for the augmented system can be expressed in the of form (\ref{52-1})
with matrices
\begin{eqnarray}
      A &=& \left[
              \begin{array}{cccc}
                -\frac{\gamma_1}{2} & \omega_p & \frac{\sqrt{\kappa\gamma_0}}{2} & 0 \\
                -\omega_p& -\frac{\gamma_1}{2} & 0 & \frac{\sqrt{\kappa\gamma_0}}{2} \\
               \frac{ -\sqrt{\kappa\gamma_0}}{2}& 0 & -\frac{\gamma_0}{2} & \omega_a \\
                0 & \frac{-\sqrt{\kappa\gamma_0}}{2} & -\omega_a & -\frac{\gamma_0}{2} \\
              \end{array}
            \right],\nonumber\\
     B &=& {\rm diag}[-\sqrt{\gamma_1},-\sqrt{\gamma_1},-\sqrt{\gamma_0},-\sqrt{\gamma_0}], \nonumber\\
     C&=& {\rm diag}[\sqrt{\gamma_1},~0,~0,~0],\nonumber\\
         D&=& [1,~0,~0,~0].\nonumber
    \end{eqnarray}
The corresponding operators $x$ and $w$ in this example are the quadrature
representations of the operators of the principal and ancillary systems and
the probing and quantum white noise processes, respectively:
\begin{eqnarray*}
x(t)&=&[ q_p(t),p_p(t), q_a(t),p_a(t)]^T\\
{\rm d}w(t)&=&[{\rm d}v_q(t),{\rm d}v_p(t),{\rm d}\bar v_q(t),{\rm d}\bar
v_p(t)]^T.
\end{eqnarray*}
Note that $y(t)$ is the position quadrature of the output of the probing
field process.

\paragraph*{Whitening filter estimates vs the mean of the principal system}
Assume the initial state $\rho_0$ of the augmented system is
Gaussian~\cite{olivares} and thus, the mean of the system operators $ m(t)=\langle  x(t)\rangle$  satisfying $\dot{ m}(t)=A' m(t)$
can serve as an estimate of the system dynamics; here the quantum
expectation is with respect to the initial quantum state $\rho_0$,
$\langle\cdot\rangle={\rm tr}[\cdot \rho_0]$. On the other hand,  the
estimates conditioned on the homodyne detection data are generated by the
quantum Kalman filter (\ref{54}). We now compare these estimates.

\begin{figure}
  \includegraphics[width=8.5cm]{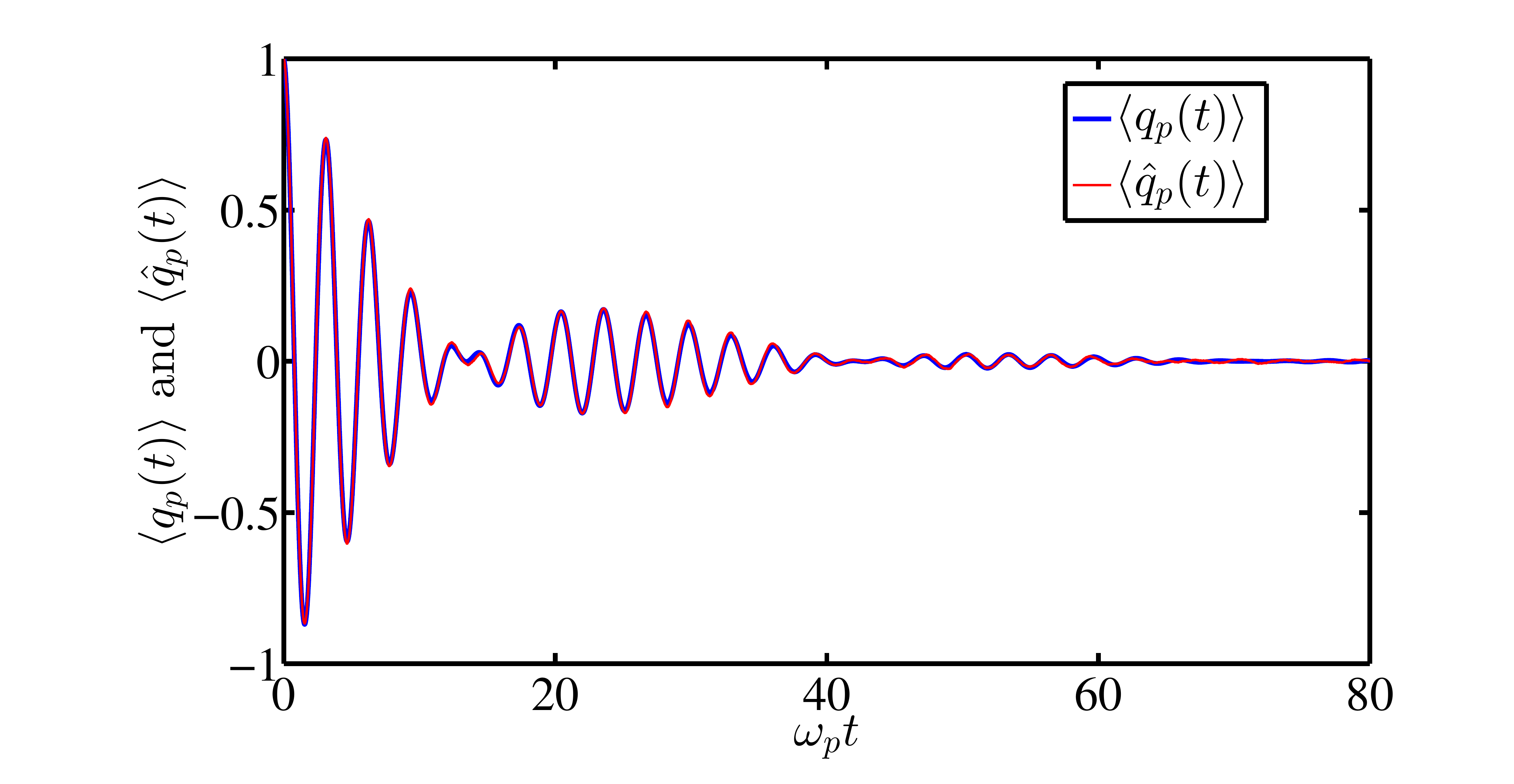}\\
  \caption{Unconditional and conditional means of the position components for the principal system with $\kappa=2{\rm GHz}$, $\omega_p=\omega_a=10{\rm GHz}$, $\gamma_0=0.6{\rm GHz}$, and $\gamma_1=0.8{\rm GHz}$.}\label{filtering4}
\end{figure}

We choose the parameters of the system as $\omega_p=\omega_a=10{\rm GHz}$,
$\kappa=2{\rm GHz}$, $\gamma_0=0.6{\rm GHz}$, and $\gamma_1=0.8{\rm GHz}$
and assume that the initial mean $ m(0)=[1,0,0,0]^T$ of the unconditional
dynamical variables is the same as the initial conditional expectation for
the quantum Kalman filter, i.e., $\hat x(0)=m(0)$ where the first element
of $\hat x(0)$ is $\langle\hat q_p(0)\rangle=1$. In Fig.~\ref{filtering4},
the red line is the trajectory of the mean of the unconditional position
operator $q_p(t)$ for the principal system. The oscillations of the curve
envelopes are caused by the disturbance of the ancillary system, which
indicates that energy is exchanged between the principal and the ancillary
system showing non-Markovian characteristics. Compared with the
unconditional trajectory, the blue line in Fig.~\ref{filtering4} shows the
average trajectory of the conditional expectation of the position $\langle
\hat q_p(t)\rangle$ obtained by averaging over 10000
realizations. It matches the red line very closely. This shows that on
average, the whitening quantum filter estimates dynamics of the unconditional
variable of the principal system.

\paragraph*{The spectrum with respect to the white noise field}

Generally, the output field spectrum is indicative of properties of the
system. To calculate the power spectral density of the output of the
probing field, we recall the quantum stochastic differential equation of
the augmented system
\begin{eqnarray}\label{55}
\left[
        \begin{array}{c}
          {\rm d} d_p(t)\\
        {\rm d} { a}_0(t)\\
        \end{array}
      \right]&=&\left[
                \begin{array}{cc}
                  -{\rm i}\omega_p-\frac{\gamma_1}{2} & \frac{\sqrt{\kappa\gamma_0}}{2} \\
                  -\frac{\sqrt{\kappa\gamma_0}}{2} & -{\rm i}\omega_a-\frac{\gamma_0}{2}\\
                \end{array}
              \right]\left[
        \begin{array}{c}
          d_p(t)\\
          a_0(t)\\
        \end{array}
      \right]{\rm d}t\nonumber\\
      &&~~~~~~~~~~~~~~-\left[
                                  \begin{array}{cc}
                                     \sqrt{\gamma_1} & 0 \\
                                    0 &  \sqrt{\gamma_0} \\
                                  \end{array}
                                \right]
      \left[
        \begin{array}{c}
          {\rm d}B_p(t)\\
         {\rm d}B_a(t)\\
        \end{array}
      \right],
\end{eqnarray}
where ${\rm d}B_p(t)$ and ${\rm d}B_a(t)$ are the probing field and ancillary white noise field processes, respectively.
The output equation with respect to the probing field ${\rm d}B_p(t)$ is
\begin{equation}\label{56}
  {\rm d}B_{\rm out}(t)=\sqrt{\gamma_1}d_p(t){\rm d}t+{\rm d}B_p(t).
\end{equation}

By detecting the position quadrature of the output field ${\rm
  d}y_{q}(t)=\frac{1}{\sqrt{2}}[ {\rm d}B_{\rm out}(t)+ {\rm d}B_{\rm
  out}^\dagger(t)]$, the power spectral density of the output field can be
calculated to be
\begin{equation}\label{59}
  S(\tilde \omega)=\frac{1}{2}(|G_1(\mathrm{i}\tilde \omega)|^2+|G_2(\mathrm{i}\tilde \omega)|^2)
\end{equation}
where $\tilde \omega=\omega_p-\omega$, and $G_1$ and $G_2$ are the transfer
functions from the probing field process and the quantum white noise process
to the output field process, respectively. The square of their norms are
expressed as
\begin{eqnarray}
  |G_1(\mathrm{i}\tilde \omega)|^2 &=&\frac{\Upsilon(\tilde
    \omega)+\frac{1}{16}(\kappa-\gamma_1)^2\gamma_0^2}{\Upsilon(\tilde
    \omega)+\frac{1}{16}(\kappa+\gamma_1)^2\gamma_0^2}, \nonumber\\
 |G_2(\mathrm{i}\tilde \omega)|^2 &=&
 \frac{\frac{\kappa\gamma_1\gamma_0^2}{4}}{\Upsilon(\tilde
   \omega)+\frac{1}{16}(\kappa+\gamma_1)^2\gamma_0^2}
\end{eqnarray}
with $\Upsilon(\tilde \omega)=(\frac{{\gamma_1}^2}{4}+\tilde
\omega^2)(\tilde \omega-\Delta)^2+\frac{\gamma_0^2\tilde
  \omega^2}{4}-\frac{\kappa\gamma_0\tilde \omega}{2}(\tilde
\omega-\Delta)$, $\Delta=\omega_p-\omega_a$.

Although the total output spectrum given in (\ref{59}) is flat  due to
the passivity properties of the system~\cite{guofengzhang2013}, we can
apply a coherent probing field whose strength is much higher than the
strength of the ancillary quantum white noise and thus the spectrum
$|G_1(\tilde \omega)|^2$ can be observed. This allows us to calculate the
spectrum $|G_2(\tilde \omega)|^2$ which reflects the influence of the ancillary
system on the output field.

\begin{figure*}
 \subfigure[$|G_2(\tilde \omega)|^2$ varying with $\kappa$ where $\Delta=0, \gamma_0=0.1{\rm GHz}$, and $\gamma_1=0.8{\rm GHz}$.]{
 \label{fig:subfig:a}
    \includegraphics[width=5.6cm]{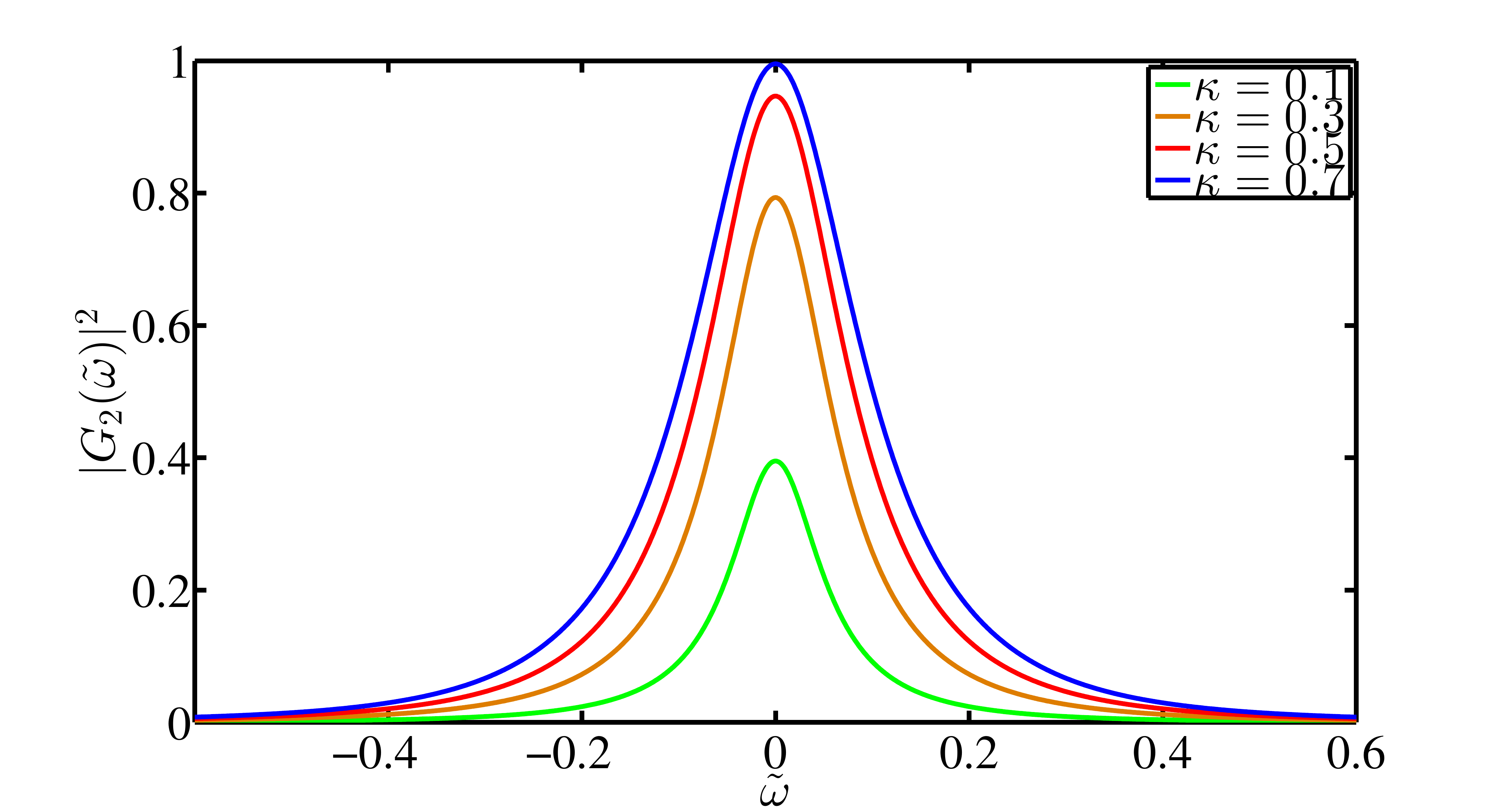}}
\subfigure[$|G_2(\tilde \omega)|^2$ varying with $\Delta$ where $\kappa=0.1{\rm GHz}, \gamma_0=0.1{\rm GHz}$, and $\gamma_1=0.8{\rm GHz}$.]{
    \label{fig:subfig:b} 
    \includegraphics[width=5.6cm]{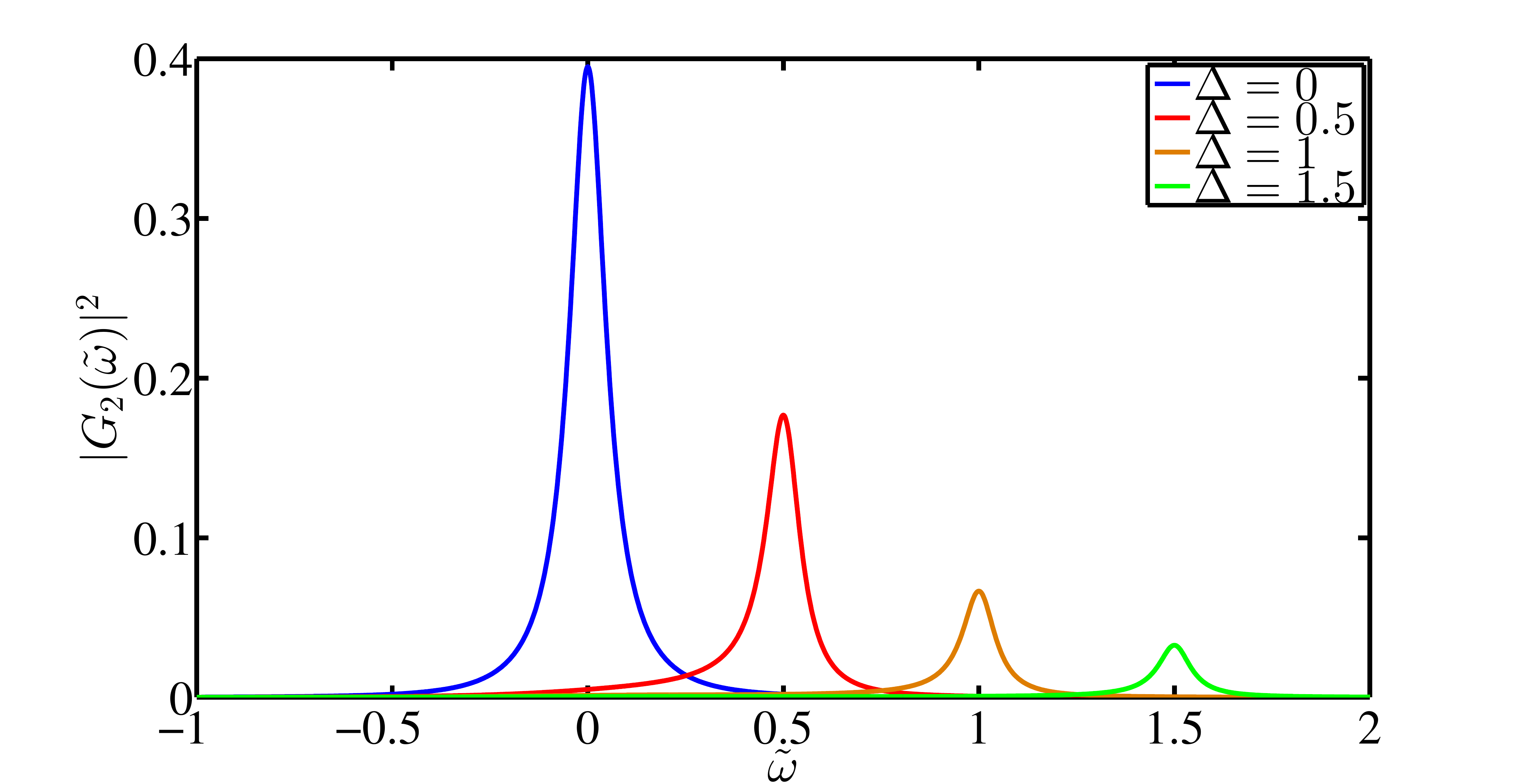}}
  \subfigure[$|G_2(\tilde \omega)|^2$ varying with $\gamma_0$ where $\Delta=0, \kappa=0.1{\rm GHz}$, and $\gamma_1=0.8{\rm GHz}$.]{
    \label{fig:subfig:c} 
    \includegraphics[width=5.6cm]{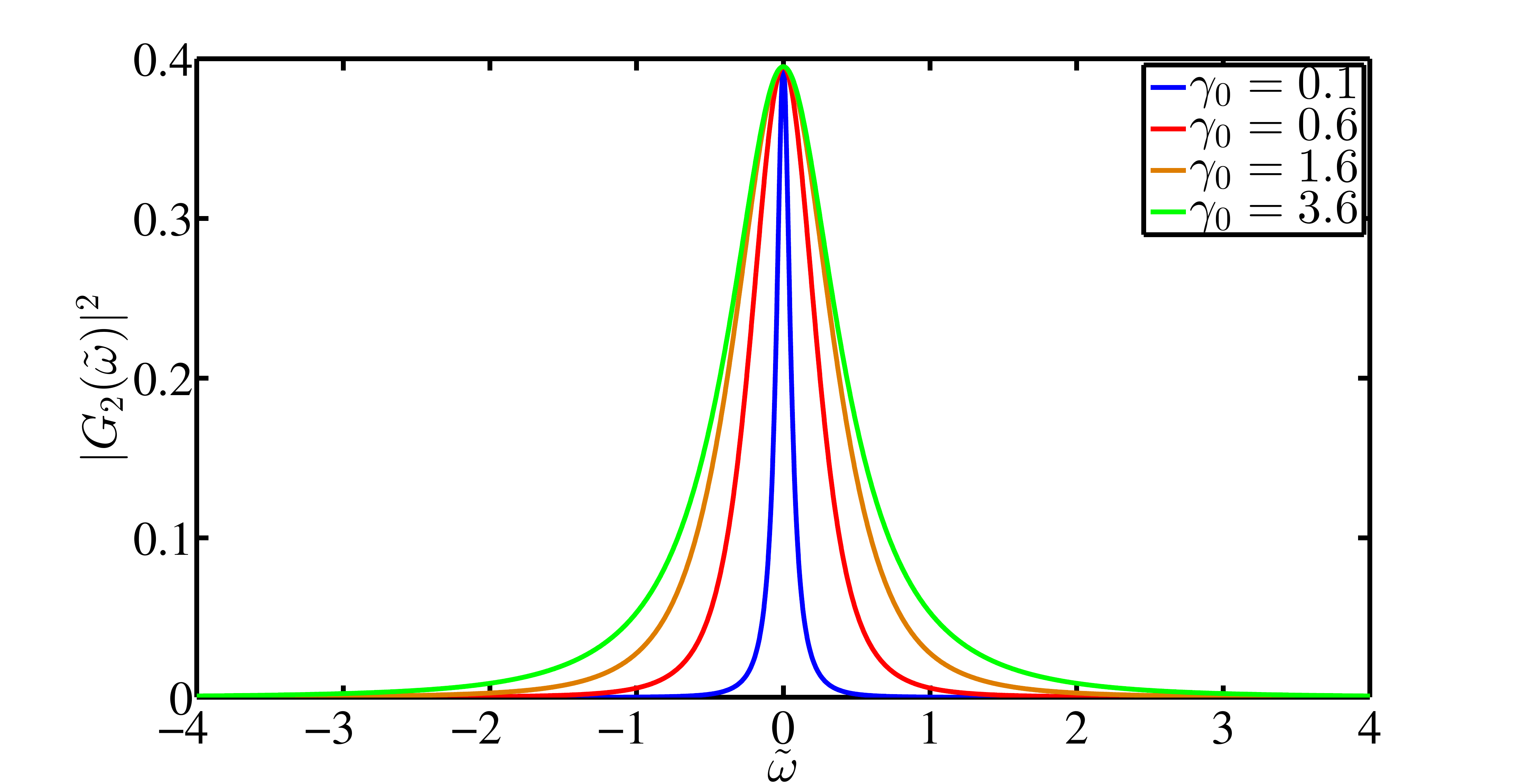}}
  \caption{The variation of the spectrum $|G_2(\tilde \omega)|^2$  versus varying parameters.}
\end{figure*}

Fig.~\ref{fig:subfig:a} shows the power spectral density $|G_2(\tilde
\omega)|^2$ varying with the coupling strength $\kappa$. Here, we assume
that there is no detuning (i.e., $\Delta=0$) and the damping rates of the
principal system to the probing field and of the ancillary system to the
quantum white noise field are $\gamma_1=0.8{\rm GHz}$  and
$\gamma_0=0.1{\rm GHz}$. As the coupling strength $\kappa$ is increased,
the amplitude of the noise spectrum is increased, which means that the
disturbance for the principal system becomes stronger.


The power spectral density $|G_2(\tilde \omega)|^2$ varying with the
detuning $\Delta$ is plotted in Fig.~\ref{fig:subfig:b} with parameters
$\kappa=0.1{\rm GHz}, \gamma_0=0.1{\rm GHz}$, and $\gamma_1=0.8{\rm
  GHz}$. When there is no detuning, the noise is
strong at the system frequency, as the blue line shows. As the detuning is
increased via decreasing the angular frequency of the ancillary system, the
spectrum $|G_2(\tilde \omega)|^2$ is driven away from the system frequency,
and its amplitude is reduced as well. This illustrates that the
non-Markovian effect generated by the ancillary system becomes weaker as
the detuning is increased. When
the detuning is large enough, the dynamics of the ancillary system become
negligible. This is consistent with the results in~\cite{Doherty1999}.

Fig.~\ref{fig:subfig:c} shows the power spectral density $|G_2(\tilde
\omega)|^2$ varying with the damping rate $\gamma_0$. As predicted, the
bandwidth of the Lorentzian spectrum is broader as $\gamma_0$ increases.

 \subsection{Quantum filter for a non-Markovian single qubit system}
A qubit is a basic unit of quantum information, and is defined on a two-dimensional complex Hilbert space $\mathfrak{h}_q$ spanned by the Pauli matrices~\cite{nilsen}
\begin{equation}\label{61}
  \sigma_x=\left[
             \begin{array}{cc}
               0& 1 \\
               1 & 0 \\
             \end{array}
           \right], \quad \sigma_y=\left[
             \begin{array}{cc}
               0& -{\rm i} \\
               {\rm i} & 0 \\
             \end{array}
           \right], \quad  \sigma_z=\left[
             \begin{array}{cc}
               1& 0 \\
               0 & -1 \\
             \end{array}
           \right].
\end{equation}



In addition, the ladder operators for the qubit system
\begin{equation}\label{62}
  \sigma_-=\left[
             \begin{array}{cc}
               0& 0 \\
              1 & 0 \\
             \end{array}
           \right], \sigma_+=\left[
             \begin{array}{cc}
               0& 1\\
               0 & 0 \\
             \end{array}
           \right]
\end{equation}
are utilized to describe a state flip between the ground state $|0\rangle$ and the excited state $|1\rangle$.
The ladder operators can also be used to describe the interaction with external systems, e.g., in the Jaynes-Cummings model~\cite{Dwalls}.

The Hamiltonian of the single qubit system we consider is given as
\begin{equation}\label{63}
  \mathsf{H}_{q}=\frac{\omega_q}{2}\sigma_z,
\end{equation}
where $\omega_q$ is the qubit working frequency characterizing the energy difference between the ground state and and the excited state.

\begin{figure}
  \centering
  \includegraphics[width=8.5cm]{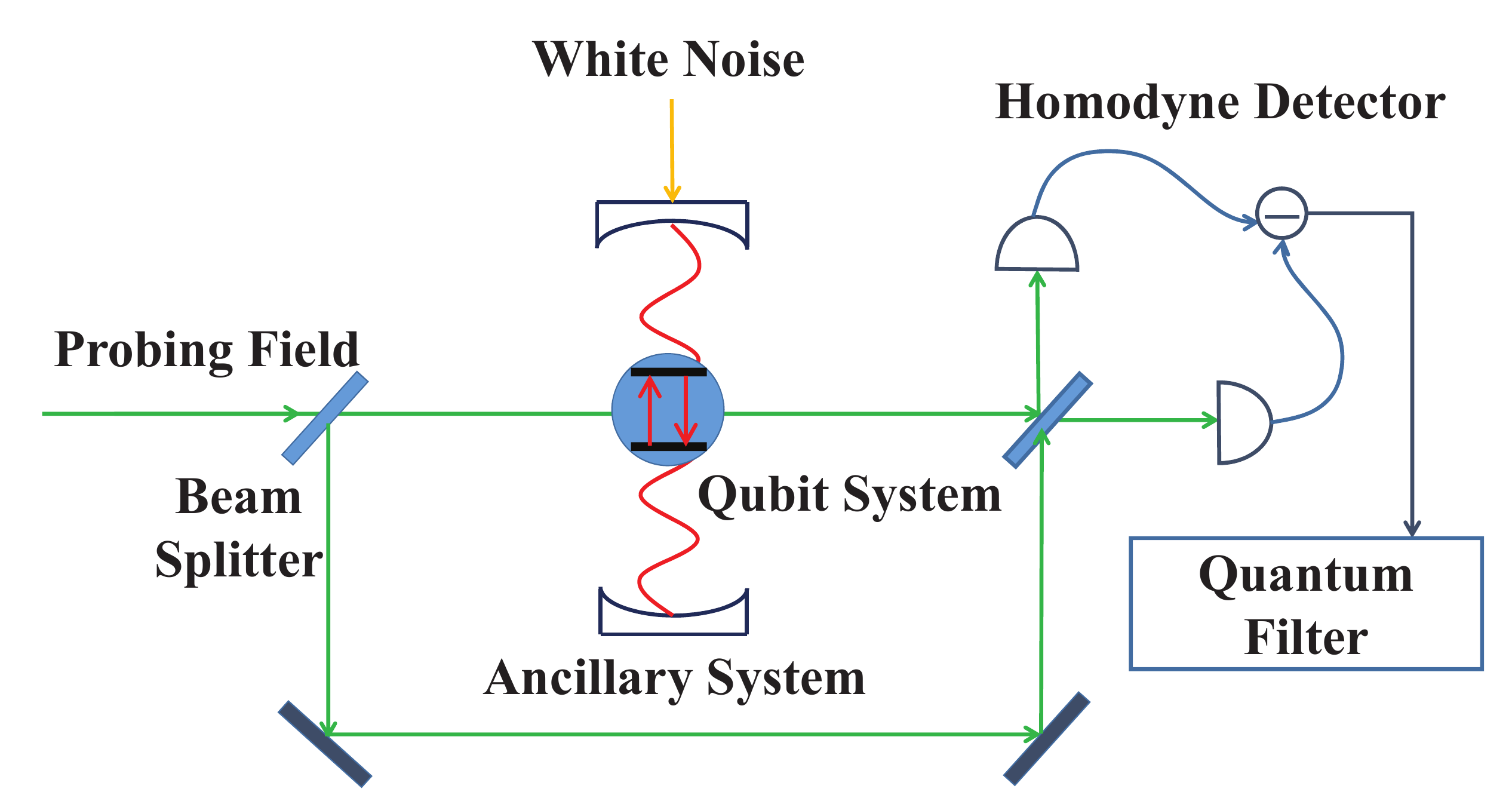}\\
  \caption{Illustrative example of a non-Markovian qubit system disturbed by Lorentzian noise.}\label{example}
\end{figure}
A non-Markovian qubit system can also be represented as an augmented system
(\ref{10}), where the Hamiltonian of the principal system is replaced by
$\mathsf{H}_{q}$ and the coupling operator $L_p$ and the direct coupling
operator $z$ are defined on the Hilbert space $\mathfrak{h}_q$. Thus, we
can write down a quantum filter for the qubit system in the Heisenberg
picture as Eq.~(\ref{43-7-2}). However, these filter equations are infinite
dimensional. Hence, for a non-Markovian qubit system, it is useful to
calculate the evolution of the unconditional and conditional states in the
Schr{\"o}dinger picture using the master equation (\ref{16}) and the
stochastic master equation (\ref{43-7-3}), respectively.


As an example, consider a single qubit system disturbed by Lorentzian noise
arising from an ancillary system as shown in Fig.~\ref{example}. As
in the previous section, the ancillary system is a single mode linear
quantum system with the Hamiltonian $\omega_a a^\dagger_0 a_0$, the
coupling operator $L_a=\sqrt{\gamma_0}a_0$, and the fictitious output
$c=-\frac{\sqrt{\gamma_0}}{2}a_0$. Here, the damping rate of the ancillary
system with respect to the quantum white noise field is $\gamma_0=0.6{\rm
  GHz}$. In addition, the qubit is coupled with the ancillary system and
the probing field through the direct operator $z=\sqrt{\kappa_1}\sigma_y$
and the coupling operator $L_q=\sqrt{\gamma_q}\sigma_x$, respectively, with
the direct coupling strength $\kappa_1=1{\rm GHz}$ and the damping rate of
the qubit $\gamma_q=0.8{\rm GHz}$., The single qubit system is initialized
in a  state $\rho_q(0)=\frac{1}{2}({\rm I}+\sigma_x)$, where $\rho_q$ is
the density matrix of the qubit and ${\rm I}$ is the $2\times2$ identity
matrix.
The angular frequency of the ancillary system is equal to that of the
qubit,  $\omega_0=\omega_q=10{\rm GHz}$.
Note that the dynamics of the ancillary systems cannot be eliminated via the adiabatic elimination which is only valid for the off-resonant case, i.e., when there exists a large detuning  between the qubit system and the ancillary systems~\cite{Doherty1999}.

\begin{figure}
  \includegraphics[width=8.5cm]{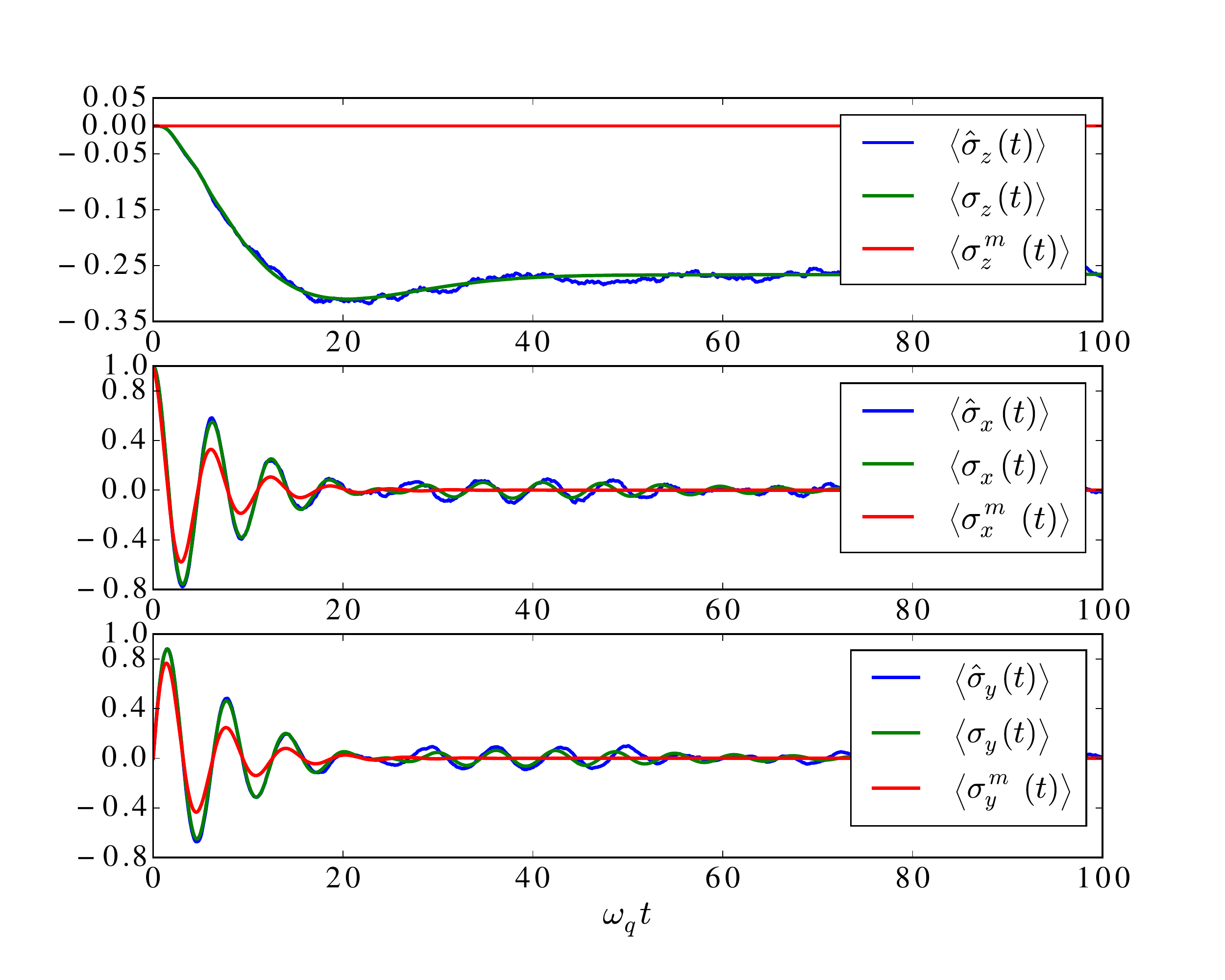}\\
  \caption{The observables for the qubit system in both non-Markovian and Markovian cases. The unconditional and averaged conditional expectation of the observables in the non-Markovian case are denoted as $\langle\sigma_.(t)\rangle$ (green lines) and $\langle\hat\sigma_.(t)\rangle$ (blue lines), respectively. For the Markovian qubit, the unconditional expectations of the observables are denoted as $\langle\sigma_.^m(t)\rangle$ (red lines). }\label{qubit}
\end{figure}

Fig.~\ref{qubit} shows the evolution of both the unconditional and averaged
conditional expectation values of the observables $\sigma_x$, $\sigma_y$,
and $\sigma_z$ for the non-Markovian qubit system. The conditional state
$\hat \rho(t)$ for the augmented system can be obtained from the quantum
filter (\ref{43-7-3}) and thus the conditional expectation of the
observables for the qubit system can be calculated as
$\langle\hat\sigma\rangle={\rm tr}[(\sigma\otimes{\rm I})\hat \rho_t]$,
where $\sigma\in\{\sigma_x,\sigma_y,\sigma_z\}$. Here, ${\rm I}$ is the
identity matrix defined on the Hilbert space of the ancillary system. The
averages of the three conditional expectations
$\langle\hat\sigma_x\rangle$, $\langle\hat\sigma_y\rangle$ and
$\langle\hat\sigma_z\rangle$ are plotted as blue lines; they are obtained
by averaging over 500 realizations of the trajectories. The green lines
represent the unconditinal expectations $\langle\sigma_x\rangle$,
$\langle\sigma_y\rangle$ and $\langle\sigma_z\rangle$
which are obtained from the master equation for the
augmented system (\ref{16}). Once again, we observe that the whitening
quantum filter can estimate the non-Markovian evolution of the single qubit
system.

To compare with the non-Markovian trajectories, the unconditional
expectation values of the observables $\sigma_x$, $\sigma_y$, and
$\sigma_z$ for the qubit system in the Markovian case are also plotted as
the red lines in Fig~\ref{qubit}, where the qubit is directly open to the
quantum white noise field and the probing field. In this case, the system
dynamics obey a Markovian master equation
\[
\dot \rho_q(t)=-{\rm
  i}[\mathsf{H}_q,\rho_q(t)]+\mathcal{L}^*_{z}(\rho_q(t))+\mathcal{L}^*_{L_q}(\rho_q(t)).
\]
The comparison shows that not only does the qubit in the Markovian case
damp faster than
that in the non-Markovian case but also the stationary states of the qubit
in the two cases are different.

\begin{figure}
  \includegraphics[width=8.5cm]{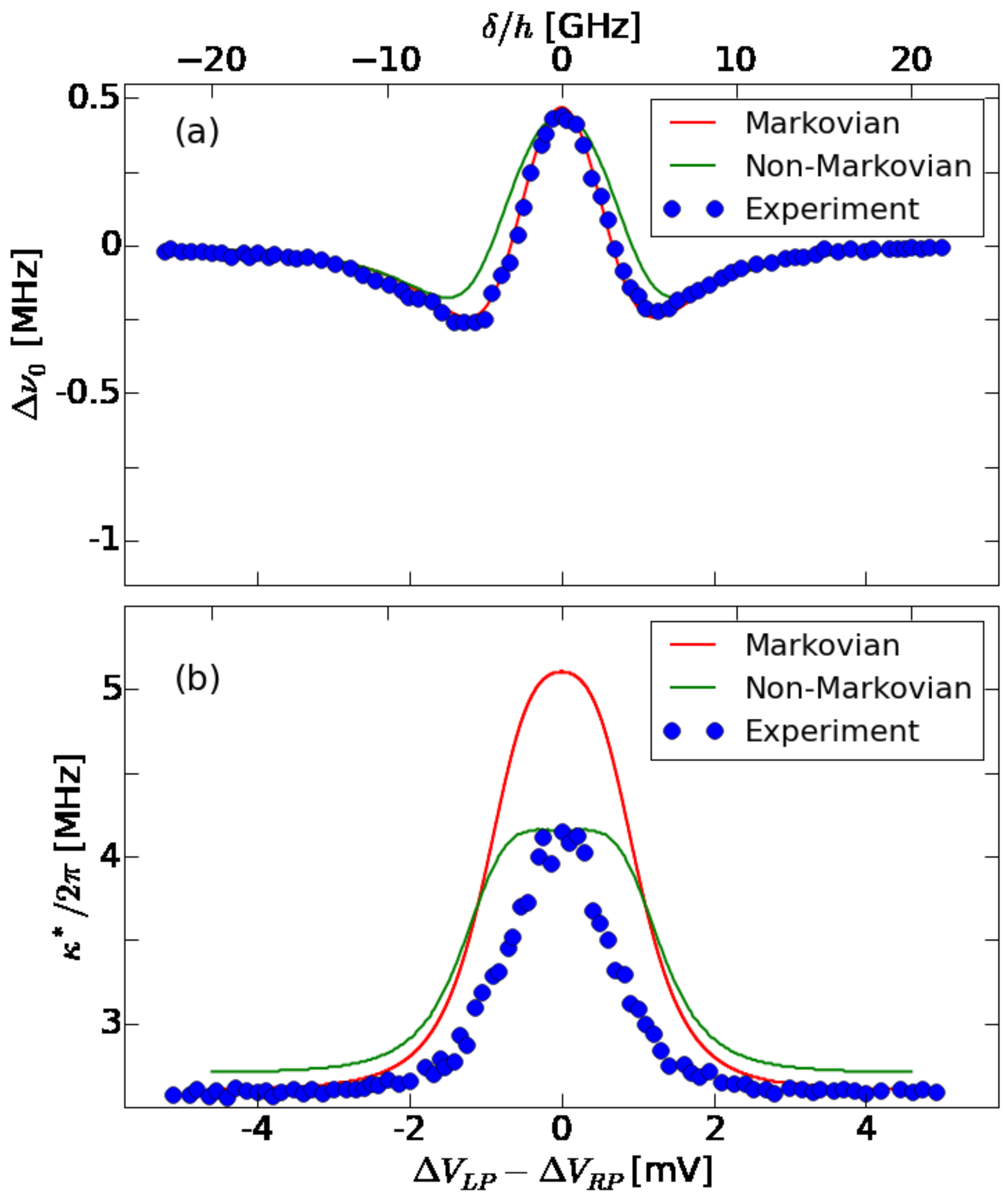}\\
  \caption{(a) Resonator frequency shift $\Delta\nu_0$ and (b) broadened resonator linewidth $\kappa^*$ for the hybrid solid-state system, where the experimental data and the curves for the Markovian system model are obtained from~\cite{Frey} and the non-Markovian curves are calculated using an augmented Markovian system model.  }\label{dotcavity}
\end{figure}

\section{Application to an experiment on a hybrid solid-state quantum device}\label{sec5}
In some experiments involving solid-state systems,  Markovian system models cannot completely explain some subtle phenomena.
For example, in the
experiment for effectively measuring a double quantum dot qubit in a
superconducting waveguide resonator~\cite{Frey}, the experimental data on
the broadened resonator linewidth under suitable parameters disagrees with
a calculation based on a Markovian system model. This discrepancy
in~\cite{Frey} was predicted to be caused by colored noise. In this
section, we assume that the discrepancies in the experiment are caused by
colored noise and apply our augmented system model to explore the effect of
the colored noise assumption.
\begin{figure}
  \includegraphics[width=8.5cm]{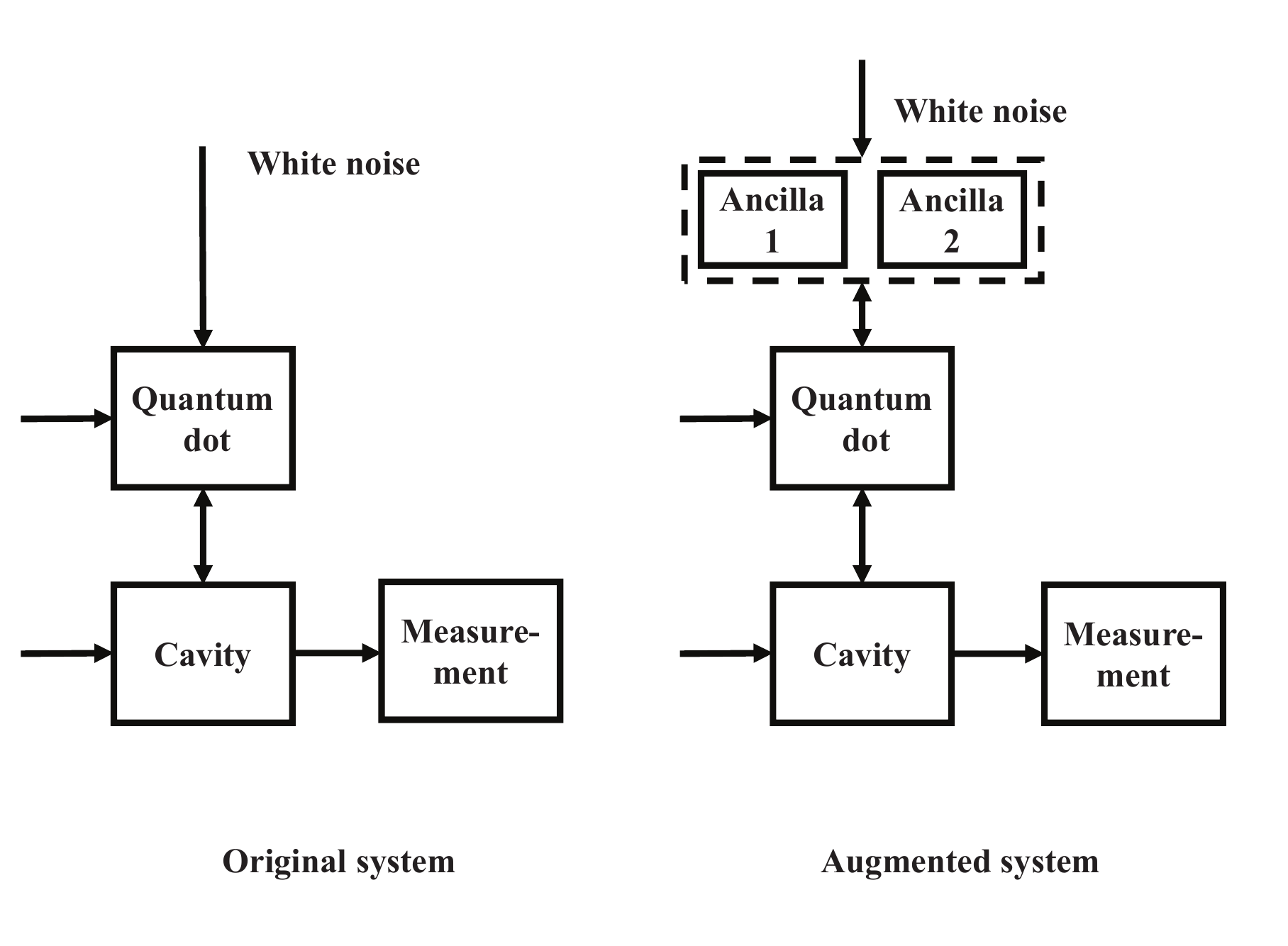}\\
  \caption{Block diagram of the hybrid solid-state system considered in~\cite{Frey}.}\label{OSAS}
\end{figure}

A block diagram of the hybrid solid-state system in~\cite{Frey} is shown in Fig.~\ref{OSAS}. In the Markovian system model considered in~\cite{Frey}, a double quantum dot qubit is directly coupled with a superconducting waveguide resonator (i.e., a cavity) which can be described by a Jaynes-Cumming Hamiltonian as
\begin{eqnarray}\label{65}
\mathsf{H}_d&=&(\nu_0-\nu_d)a_r^\dagger a_r+\frac{\nu_{qb}-\nu_d}{2}\sigma_z+g_c{\rm sin}\theta(a_r^\dagger\sigma_-+\sigma_+a_r)\nonumber\\
&&+\frac{\epsilon}{2}(a_r^\dagger+a_r)
\end{eqnarray}
where the angular frequency of the resonator is $\nu_0=6.755{\rm GHz}$  and
the frequency of the qubit is $\nu_{qb}=\sqrt{4\mu^2+\delta^2}$ with
$\mu=4.5{\rm GHz}$. Here, $\delta$ is the detuning frequency between the
qubit states, which can be varied when measuring the resonator. The
coupling strength $g_c=0.05\times 2\pi{\rm GHz}$ between the resonator and
the qubit can be modulated via a sine function ${\rm
  sin}\theta=\frac{2\mu}{\sqrt{4\mu^2+\delta^2}}$. The annihilation and
creation operators for the resonator are denoted as $a_r$ and
$a_r^\dagger$, respectively. The symbols $\nu_d$ and $\epsilon$ represent
the frequency and amplitude of the driving field, which is a weak and fixed
strength field; see~\cite{FreyThesis} for more details.

Furthermore, the qubit is coupled with one dissipative channel and one
dephasing channel which are characterized by coupling operators
$\sqrt{\bar\gamma_-}\sigma_-$ and $\sqrt{\frac{\bar\gamma_z}{2}}\sigma_z$
where
$\bar\gamma_-=\sin^2\theta\bar\gamma'_-+\cos^2\theta\bar\gamma'_z$ and $\bar\gamma_z=\cos^2\theta\bar\gamma'_-+\sin^2\theta\bar\gamma'_z$
with $\bar\gamma'_-=3.3\times2\pi{\rm GHz}$ and
$\bar\gamma'_z=100\times2\pi{\rm MHz}$. In addition, the cavity is probed
by a field; the corresponding coupling operator is $\sqrt{\bar\kappa}a_r$
with $\bar\kappa=2.6\times2\pi{\rm MHz}$. Hence, the dynamics of this
hybrid system obey the master equation
\begin{eqnarray}\label{66}
  \dot{\bar\rho}(t)&=&-{\rm i}[\mathsf{H}_d,\bar\rho(t)]+\mathcal{L}^*_{\sqrt{\bar\kappa}a_r}(\bar\rho(t))\nonumber\\
  &&~~~~~~~~+\mathcal{L}^*_{\sqrt{\bar\gamma_-}\sigma_-}(\bar\rho(t))+\mathcal{L}^*_{\sqrt{\frac{\bar\gamma_z}{2}}\sigma_z}(\bar\rho(t)),
\end{eqnarray}
where $\bar\rho(t)$ is the density matrix for the resonator and qubit system.

With this model, the transmission amplitude through the cavity is
calculated as $\langle a_r\rangle={\rm tr}[a_r\bar\rho_s]$ where
$\bar\rho_s$ represents the steady state of Eq.~(\ref{66}). The shift of
the resonance frequency $\Delta\nu_0$ and the broadened linewidth of the
cavity $\kappa^*$ can be obtained from the square of the transmission
amplitude, which are plotted as red lines in Fig.~\ref{dotcavity} (a) and
(b), respectively. Compared to the experimental data plotted as blue dots,
the shift of the resonance frequency curve in Fig.~\ref{dotcavity} (a) is
in good agreement but the broadened linewidth curve in Fig.~\ref{dotcavity}
(b) has a large discrepancy. This discrepancy is potentially caused by
colored noise as conjectured in~\cite{Frey}.

To explore the reason for the discrepancy in the broadened linewidth curves and obtain better curve matching, an augmented system model is utilized to describe the system dynamics. We assume that the dissipative channel of the quantum dot in the original system is a colored noise channel.
However, since the spectrum of the colored noise is \textit{unknown} for
this hybrid solid-state system, it is difficult to realize the ancillary
system in the augmented system by using the spectral factorization
method. From Corollary~\ref{coro}, we have known that Lorentzian noise can
be generated using a single-mode linear quantum system. When a suitable
number of single-mode linear ancillary systems are coupled with the quantum
dot through an identical operator of the quantum dot, the Lorentzian noises
generated by the ancillary systems can be combined to approximate an arbitrary
colored noise. We have tried several such approximations; the results of
these trials are given in the appendix.

One case we considered is where only one ancillary system is coupled to the
quantum dot. When the ancillary system is resonant or off-resonant to the
cavity, the peak values of the broadened resonator linewidth $\kappa^*$
can be modified to provide a good fit to the experimental data. However,
the peak values of the resonator frequency shift $\Delta\nu_0$ vary in an
opposite direction.

Next, we considered the situation where the quantum dot is coupled to two
ancillary systems, one of them was a resonant system and another one was an
off-resonant ancillary system. The block diagram for the model with two
ancillary systems is shown in Fig.~\ref{OSAS}. This modified model with
refined parameters can be described as follows. The Hamiltonian of this
modified model can be written as
\begin{equation}\label{67}
  \mathsf{H}'_d=\mathsf{H}_d+\sum_{j=1,2}({\rm i}\frac{\sqrt{\bar\kappa_j\bar\gamma_j}}{2}(\sigma_+a_j-a_j^\dagger\sigma_-)+(\nu_j-\nu_d)a_j^\dagger a_j),
\end{equation}
where $a_1$, $a_2$ and $a_1^\dagger$, $a_2^\dagger$ denote the
annihilation and creation operators for the ancillary systems, and
the respective angular frequencies of the two ancillary systems are
$\nu_1=6.755{\rm GHz}$ and  $\nu_2=1.2{\rm GHz}$. The
coupling strengths between the ancillary systems and the quantum dot
are $\bar\kappa_1=0.5\bar\gamma_-$ and $\bar\kappa_2=1.125\bar\gamma_-$,
respectively. The coupling operators of the ancillary systems with respect
to quantum white noise fields are $\sqrt{\bar\gamma_1}a_1$ and
$\sqrt{\bar\gamma_2}a_2 $ with $\bar\gamma_1=35{\rm MHz}$ and
$\bar\gamma_2=20{\rm GHz}$, respectively.
Thus, the dynamics of the augmented system can also described by a master equation
\begin{eqnarray}\label{68}
  \dot{\bar\rho}'(t)&=&-{\rm i}[\mathsf{H}'_d,\bar\rho'(t)]+\mathcal{L}^*_{\sqrt{\bar\kappa}a_r}(\bar\rho'(t))\nonumber\\
  &&~~~+\sum_{j=1,2}\mathcal{L}^*_{\sqrt{\bar\gamma_j}a_j}(\bar\rho'(t))+\mathcal{L}^*_{\sqrt{\frac{\bar\gamma_z}{2}}\sigma_z}(\bar\rho'(t)),
\end{eqnarray}
where $\bar\rho'(t)$ is the density matrix of the augmented system. The
transmission amplitude through the cavity is now $\langle a_r\rangle'={\rm
  tr}[a_r\bar\rho'_s]$, where $\bar\rho'_s$ is the steady state of
Eq.~(\ref{68}), and thus with the modified model the shift of the resonance
frequency and the broadened linewidth of the cavity can be
obtained as well.
\begin{figure}
  \includegraphics[width=8.5cm]{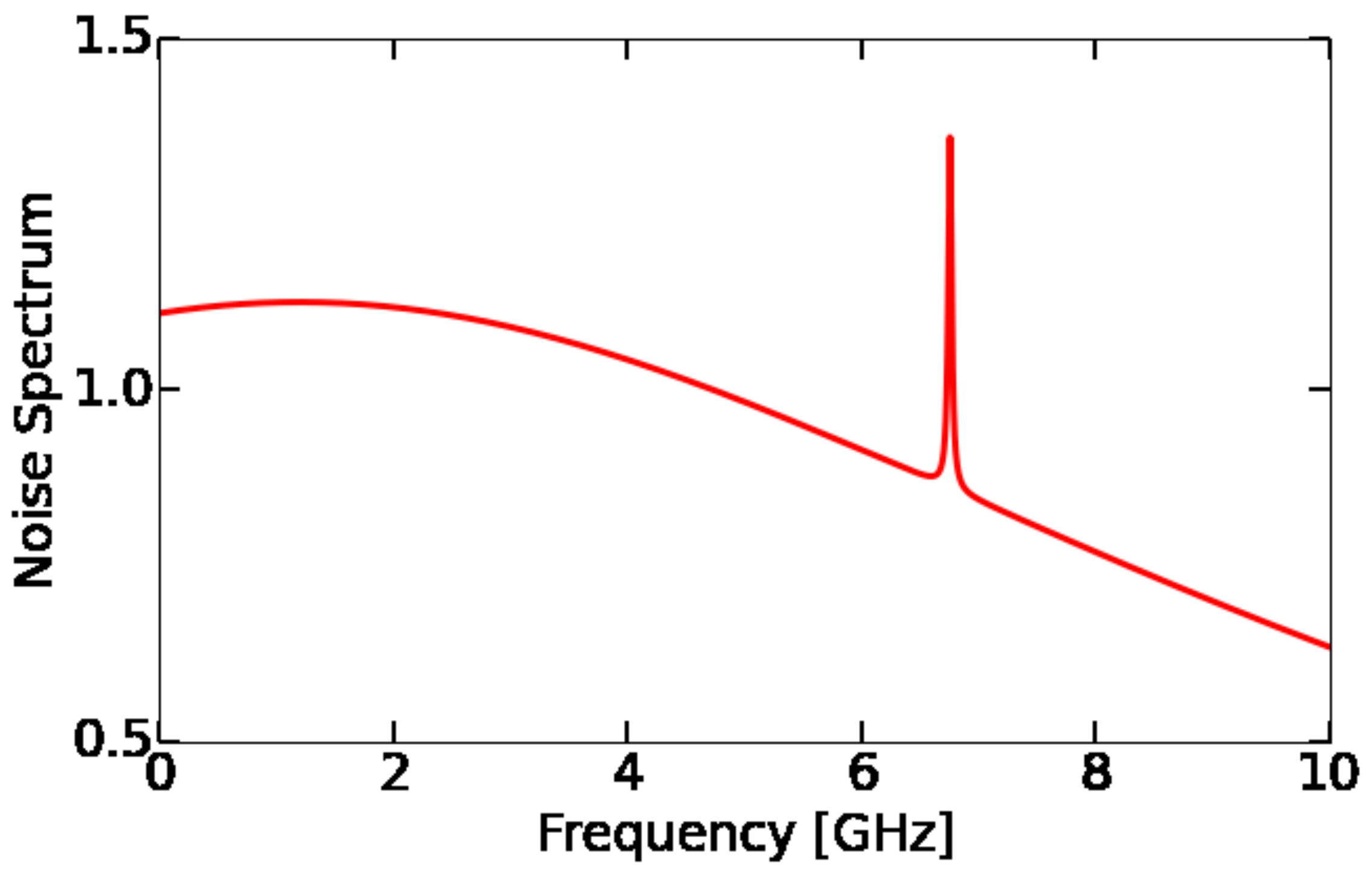}\\
  \caption{Colored noise spectrum generated by the two ancillary systems.}\label{TL}
\end{figure}

After several trials using the two-ancillary model, we obtained a broadened
linewidth which matches the experimental data significantly closer then the
Markovian model; see the green lines in Fig.~\ref{dotcavity}.  At the same
time, the curve for the shift of the
resonance frequency deviates only slightly from the experimental data.
The spectrum of the colored noise generated by the two ancillary systems is
shown in Fig.~\ref{TL}, it has a double Lorentzian shape. One
Lorentzian spectrum is sharp with an identical frequency to the cavity and
the other one is very broad with a center frequency far away from that of
both the resonator and the quantum dot. These results indicate that the
discrepancies in~\cite{Frey} can indeed be attributed to colored
noise. Possibly even better results could be achieved if more
ancillary systems were utilized. However, the computation time
increases considerably in this case due to the dimensions of the augmented
system.

\section{Conclusions}\label{sec6}
In this paper, we have presented an augmented Markovian system model for
non-Markovian quantum systems. Also, a spectral factorization approach has
been used to systematically represent a non-Markovian environment by
means of linear ancillary quantum systems. Importantly, direct
interactions between the ancillary and principal systems are introduced,
which result in non-Markovian dynamics of the principal system. Next, we
demonstrated that using this augmented system model, a whitening quantum
filter can be constructed for continuously estimating dynamics of
non-Markovian quantum systems. Such a filter has been derived for both
linear and qubit principal systems. The proposed augmented Markovian system
model has also been applied to explore the structure of unknown colored
noises in the experiment for the hybrid resonator and quantum dot system.

The augmented Markovian system model is defined on an extended Hilbert
space whose dimension is determined by the number of the linear
ancillary systems. When the dimension of the extended Hilbert space is
large, the calculation speed for the whitening quantum filter may be
slow. Hence, for future studies, it is worthwhile to explore more efficient
techniques for representing colored noise by means of ancillary systems
which can improve the computational speed of the whitening quantum filter.



\bibliographystyle{plain}        

\appendix
\section{Additional simulation results for the hybrid solid-state system}
\begin{figure}
  \includegraphics[width=8.5cm]{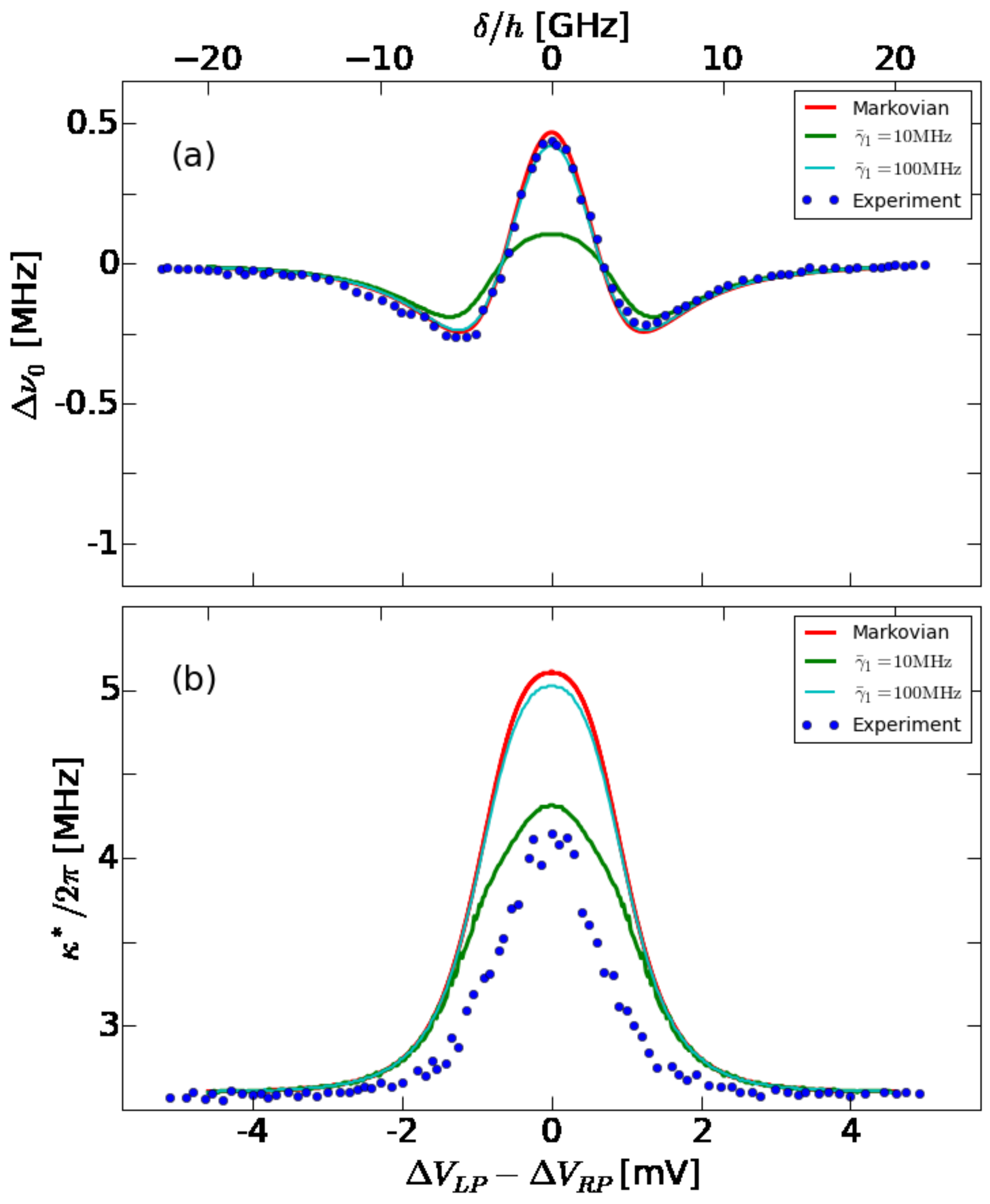}\\
  \caption{(a) Resonator frequency shift $\Delta\nu_0$ and (b) broadened resonator linewidth $\kappa^*$ for the hybrid solid-state system, where the quantum dot is coupled with one ancillary system with a resonant frequency, i.e., $\nu_1=\nu_0$.}\label{oneanciw0}
\end{figure}

We also consider three possible cases which might happen to the hybrid solid-state system in section \ref{sec5}. The first case is that the quantum dot is distrubed by Lorentzian noise where it directly interacts with one linear ancillary system with a resonant frequency, i.e., $\nu_1=\nu_0$. The results for this case are plotted in Fig.~\ref{oneanciw0}, where the experimental data and the curves for the Markovian case are plotted as blue dots and red lines, respectively. When the damping rate of the ancillary system is $\bar\gamma_1=10{\rm MHz}$, the curve for the broadened resonator linewidth $\kappa^*$ in Fig.~\ref{oneanciw0}(b) can get closer to the experimental data. However, the peak value of the corresponding curve for the resonator frequency shift $\Delta\nu_0$ in Fig.~\ref{oneanciw0}(a) is decreased too much compared with that of the experimental data. As increasing the damping rate of the ancillary system, e.g., $\bar\gamma_1=100{\rm MHz}$, the curves approach to that of the Markovian case.

\begin{figure}
  \includegraphics[width=8.5cm]{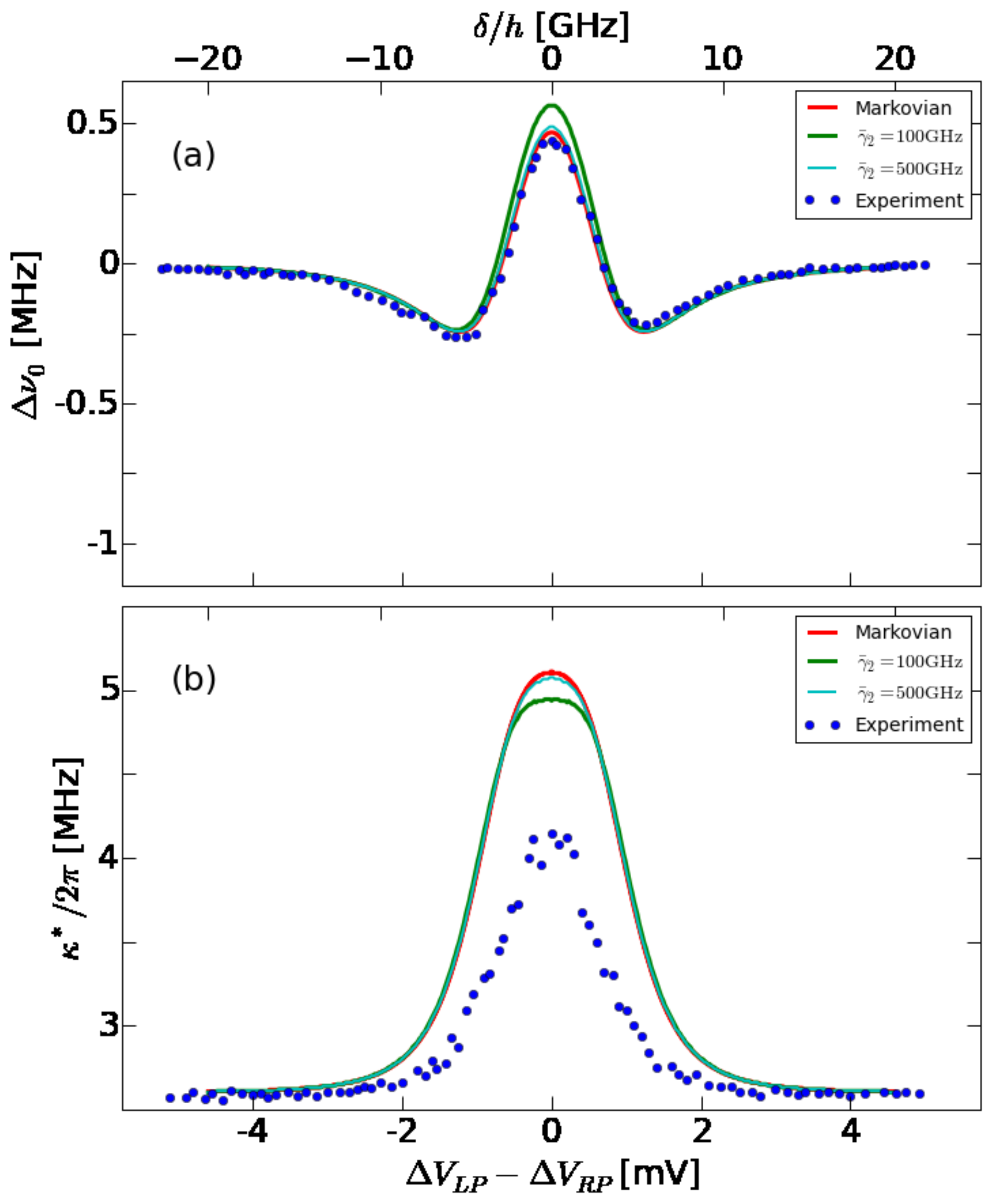}\\
  \caption{(a) Resonator frequency shift $\Delta\nu_0$ and (b) broadened resonator linewidth $\kappa^*$ for the hybrid solid-state system, where the quantum dot is coupled with one ancillary system with an off-resonant frequency, i.e., $\nu_2=1.2{\rm GHz}$.}\label{oneanci12}
\end{figure}

In the second case,  the quantum dot is coupled to one linear ancillary system with an off-resonant frequency, i.e., $\nu_2=1.2{\rm GHz}$. When the damping rate of this ancillary system is $\bar\gamma_2=100{\rm GHz}$, the peak value of the curve for the resonator frequency shift $\Delta\nu_0$ in Fig.~\ref{oneanci12}(a) is higher than that in the Markovian case. However, the corresponding peak value of the broadened resonator linewidth $\kappa^*$ in Fig.~\ref{oneanci12}(b) becomes lower than that in the Markovian case. Also, as increasing the damping rate $\bar\gamma_2$, e.g., $\bar\gamma_2=500{\rm GHz}$, both curves for $\Delta\nu_0$ and $\kappa^*$ approach to the curves in the Markovian case.

In addition, the case that the quantum dot is coupled to both one ancillary system and a Markovian dissipative channel is considered. The frequency of the ancillary system is $\nu_1=\nu_0$ and the damping rates of the ancillary system is $\bar\gamma_1=35{\rm MHz}$. The damping rate of the quantum dot with respect to the Markovian dissipative channel is kept the same as $\bar\gamma_-$. In this case, both the peak values of the resonator frequency shift $\Delta\nu_0$ and the broadened resonator linewidth $\kappa^*$ are increased.
\begin{figure}
  \includegraphics[width=8.5cm]{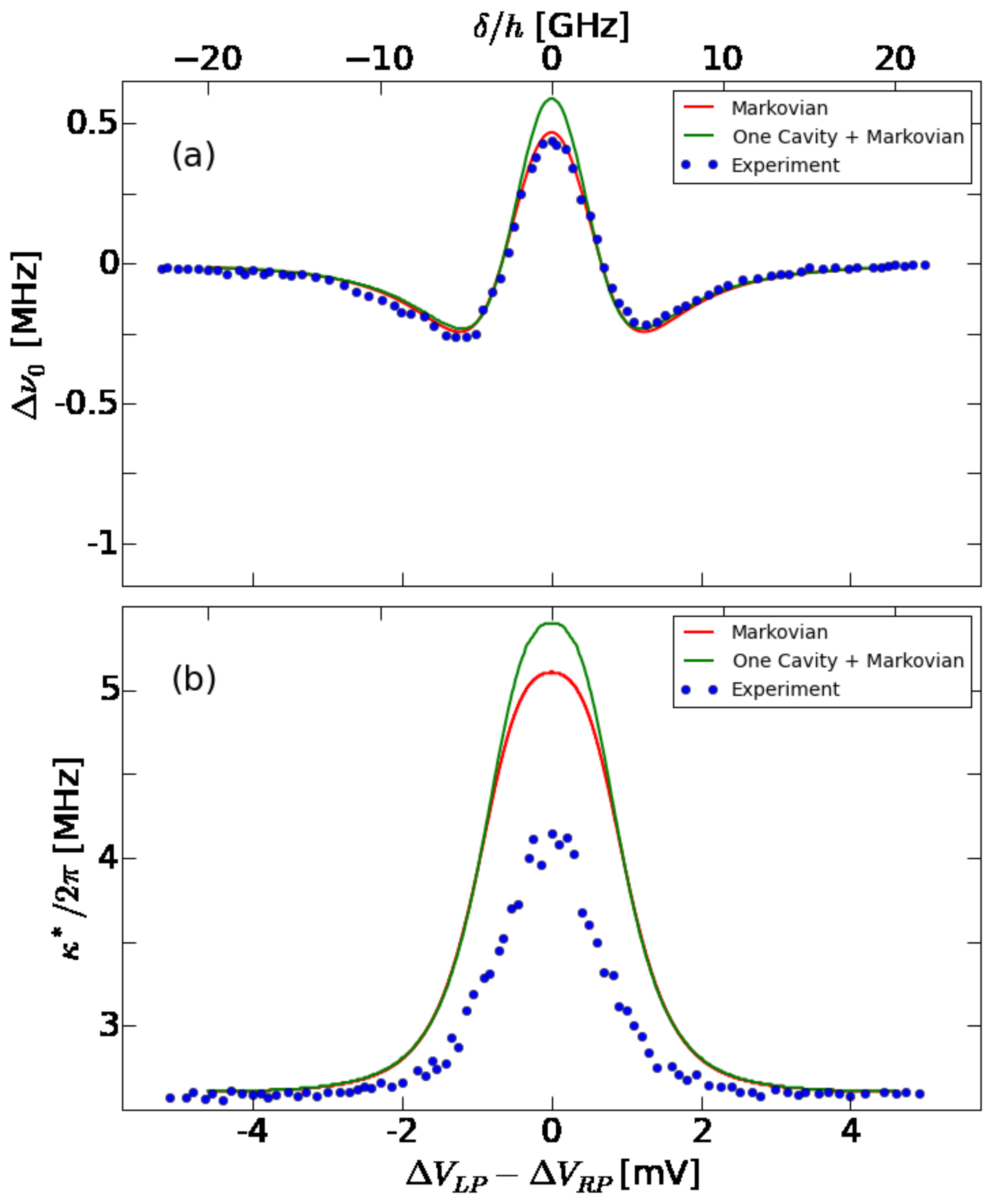}\\
  \caption{(a) Resonator frequency shift $\Delta\nu_0$ and (b) broadened resonator linewidth $\kappa^*$ for the hybrid solid-state system, where the quantum dot is coupled to both one ancillary system and a Markovian dissipative channel.}\label{oneanciMakovian}
\end{figure}

In the first two cases, the peak values of the broadened resonator linewidth $\kappa^*$  can be modified to approach to the experimental data . However, the peak values of the resonator frequency shift $\Delta\nu_0$ vary in an opposite direction. Hence, we consider to couple the quantum dot to one resonant and one off-resonant ancillary systems and thus obtain the results in section \ref{sec5}.

\end{document}